\documentclass[aps,prb,twocolumn,amsmath,amssymb,amsfonts,floatfix]{revtex4-2}

\usepackage[utf8]{inputenc}
\usepackage[english]{babel}
\usepackage[T1]{fontenc}
\usepackage{float}
\usepackage{graphicx}
\usepackage{dcolumn}
\usepackage{amsmath}
\usepackage{amssymb}
\usepackage{bm}
\usepackage[urlcolor=blue,colorlinks=true,citecolor=red,linkcolor=red,pdfstartview={FitH},bookmarks=false]{hyperref}
\usepackage{xcolor}
\usepackage{listings}
\usepackage{enumitem}
\usepackage{epstopdf}
\usepackage[normalem]{ulem}

\def\vnabla{{\boldsymbol{\nabla}}}

\def\vA{{\bf A}}
\def\vAext{{\bm{A}_{\mathrm{ext}}}}
\def\vAind{{\bm{A}_{\mathrm{ind}}}}

\def\vB{{\bf B}}

\def\vBext{{\bm{B}_{\mathrm{ext}}}}
\def\vBind{{\bm{B}_{\mathrm{ind}}}}

\def\Bind{{B_{\mathrm{ind}}}}

\def\vxhat{{\hat{\bm{x}}}}
\def\vyhat{{\hat{\bm{y}}}}
\def\vzhat{{\hat{\bm{z}}}}

\def\Phiind{{\Phi_{\mathrm{ind}}}}
\def\vj{{\bm{j}}}

\def\vjm{{\bf j}_{\mathrm{m}}}

\def\vm{{\bm{m}}}
\def\vL{{\bm{L}}}

\def\vvF{{\bf v}_{\mathrm{F}}}
\def\vF{v_{\mathrm{F}}}
\def\vFx{v_{\mathrm{F},x}}
\def\vFy{v_{\mathrm{F},y}}
\def\vp{{\bf p}}

\def\vpF{{\bf p}_{\mathrm{F}}}
\def\vpFp{{\bf p}^{\prime}_{\mathrm{F}}}
\def\vR{{\bf R}}

\def\pF{p_{\mathrm{F}}}
\def\pFx{p_{\mathrm{F},x}}
\def\pFy{p_{\mathrm{F},y}}

\def\pFxp{p^{\prime}_{\mathrm{F},x}}
\def\pFyp{p^{\prime}_{\mathrm{F},y}}
\def\EF{E_{\mathrm{F}}}
\def\NF{N_{\mathrm{F}}}
\def\FS{{\mathrm{FS}}}
\def\thetaF{{\theta_{\mathrm{F}}}}
\def\thetaFp{{\theta^{\prime}_{\mathrm{F}}}}

\def\muB{\mu_{\mathrm{B}}}
\def\kB{k_{\mathrm{B}}}

\def\Tc{T_{\mathrm{c}}}

\def\Omegac{\Omega_{\mathrm{c}}}

\def\Deltas{{\Delta_{\mathrm{s}}}}
\def\tDelta{{\tilde{\Delta}}}
\def\tDeltas{{\tilde{\Delta}_{\mathrm{s}}}}
\def\dxtyt{{{d_{x^2-y^2}}}}
\def\Dxtyt{{\Delta_{d_{x^2-y^2}}}}
\def\Dxy{{\Delta_{d_{xy}}}}
\def\etaxtyt{{\eta_{d_{x^2-y^2}}}}
\def\etaxy{{\eta_{d_{xy}}}}
\def\chixtyt{{\chi_{d_{x^2-y^2}}}}
\def\chixy{{\chi_{d_{xy}}}}
\def\Dplus{{\Delta_{+}}}
\def\Dminus{{\Delta_{-}}}
\def\Dpm{{\Delta_{\pm}}}
\def\Dmp{{\Delta_{\mp}}}

\def\Lzorb{{\hat{L}_z^{\mathrm{orb}}}}

\def\vsigma{{\boldsymbol{\sigma}}}
\def\tgamma{{\tilde{\gamma}}}
\def\tGamma{{\tilde{\Gamma}}}
\def\tgammah{{\tilde{\gamma}_{\mathrm{h}}}}
\def\tgammas{{\tilde{\gamma}_{\mathrm{s}}}}
\def\vtgammat{{\tilde{\boldsymbol{\gamma}}_{\mathrm{t}}}}
\def\gammah{{\gamma_{\mathrm{h}}}}
\def\gammas{{\gamma_{\mathrm{s}}}}
\def\vgammat{{\boldsymbol{\gamma}_{\mathrm{t}}}}
\def\tGammain{{\tilde{\Gamma}_{\mathrm{in}}}}
\def\tgammahout{{\tilde{\gamma}_{\mathrm{h,out}}}}
\def\tgammahin{{\tilde{\gamma}_{\mathrm{h,in}}}}
\def\Gammain{{\Gamma_{\mathrm{in}}}}

\def\gammahout{{\gamma_{\mathrm{h,out}}}}
\def\gammahin{{\gamma_{\mathrm{h,in}}}}
\def\gammaE{{\gamma_{\mathrm{E}}}}
\def\xiy{{\xi_y}}

\def\decayLength{{r_0}}

\def\regimeLengthXi{{20\xi_0}}

\allowdisplaybreaks

\begin{document}

\title{Enhanced chiral edge currents and orbital magnetic moment in chiral $d$-wave superconductors from mesoscopic finite-size effects}

\author{P. Holmvall}
\affiliation{Department of Physics and Astronomy,
	Uppsala University, Box 516, S-751 20, Uppsala, Sweden}
\author{A. M. Black-Schaffer}
\affiliation{Department of Physics and Astronomy,
	Uppsala University, Box 516, S-751 20, Uppsala, Sweden}

\date{\today}

\begin{abstract}
Chiral superconductors spontaneously break time-reversal symmetry and host topologically protected edge modes, supposedly generating chiral edge currents which are typically taken as a characteristic fingerprint of chiral superconductivity. However, recent studies have shown that the total edge current in two dimensions (2D) often vanishes for all chiral superconductors except for chiral $p$-wave, especially at low temperatures, thus severely impeding potential experimental verification and characterization of these superconductors.
In this work, we use quasiclassical theory of superconductivity to study mesoscopic disc-schaped chiral $d$-wave superconductors. We find that mesoscopic finite-size effects cause a dramatic enhancement of the total charge current and orbital magnetic moment (OMM), even at low temperatures. We study how these quantities scale with temperature, spontaneous Meissner screening, and system radius $\mathcal{R} \in [5,200]\xi_0$ with superconducting coherence length $\xi_0$. We find a general $1/\mathcal{R}$ scaling in the total charge current and OMM for sufficiently large systems, but this breaks down in small systems, instead producing a local maximum at $\mathcal{R} \approx 10{\text{--}}20\xi_0$ due to mesoscopic finite-size effects.
These effects also cause a spontaneous charge-current reversal opposite to the chirality below $\mathcal{R} < 10\xi_0$. Our work highlights mesoscopic systems as a route to experimentally verify chiral $d$-wave superconductivity, measurable with magnetometry.
\end{abstract}

\maketitle

\section{Introduction}
\label{sec:intro}
Chiral superconductors have recently sparked great interest \cite{Kallin:2012,Kallin:2016,Mizushima:2016,Volovik:2019}, as they are topologically non-trivial and have been suggested as a platform to realize topological quantum computing \cite{Beenakker:2013,Sato:2017,Mercado:2022,Margalit:2022,Huang:2023,Li:2023}. Chiral superconductivity, and more generally superfluidity \cite{Volovik:2003,Vollhardt:2013,Volovik:2020}, is characterized by a multi-component and complex-valued order parameter $\Delta(\vR,\vpF)$ with a ground state which breaks time-reversal symmetry in the bulk \cite{Sigrist:1991}, $\Delta(\vR,\vpF) = \Delta_1(\vpF) \pm i\Delta_2(\vpF) = |\Delta|e^{i\chi}e^{\pm i|M|\thetaF}$, where $\vR$ is the center-of-mass coordinate, $\vpF = \pF(\cos\thetaF,\sin\thetaF)$ is the Fermi momentum on the Fermi surface (FS), $|\Delta|$ is the maximum gap in the quasiparticle spectrum, and $\chi$ is the superconducting phase. Here, $M=\pm|M|$ is the Chern number of the two degenerate ground states of opposite chiralities and related to the winding of the order parameter on the FS.
Even (odd) $M$ correspond to spin-singlet (spin-triplet) superconductivity, where $|M|=1,2,3$ in two dimensions (2D) generate chiral $p,d,f$-wave, respectively.
Early studies have primarily focused on spin-triplet chiral $p$-wave or $f$-wave superconductivity \cite{Mackenzie:2003,Kallin:2012,Kallin:2016,Suh:2020,Duan:2021,Bae:2021,Hayes:2021}. Interestingly, recent studies have also proposed spin-singlet chiral $d$-wave superconductivity in many materials, including~twisted bilayer cuprates \cite{Can:2021:a,Can:2021:b}, $\textrm{SrPtAs}$ \cite{Biswas:2013,Fischer:2014,Ueki:2019,Ueki:2020}, ${\mathrm{Sn/Si}}(111)$ \cite{Ming:2023}, twisted bilayer graphene \cite{Venderbos:2018,Su:2018,Fidrysiak:2018,Xu:2018,Kennes:2018,Liu:2018,Gui:2018,Wu:2019,Fischer:2021}, ${\mathrm{Bi/Ni}}$ bilayers \cite{Gong:2017,Hosseinabadi:2019}, $\mathrm{U}{\mathrm{Ru}}_{2}{\mathrm{Si}}_{2}$ \cite{Kasahara:2007,Kasahara:2009,Shibauchi:2014,Iguchi:2021}, and $\textrm{LaPt$_3$P}$ \cite{Biswas:2021}.

The topologically non-trivial bulk order parameter of chiral superconductors leads to topologically protected chiral edge modes through the bulk-boundary correspondence \cite{Volovik:1997,Schnyder:2008,Hasan:2010,Qi:2011,Tanaka:2012,Graf:2013,Black-Schaffer:2014:b}.
The chiral edge modes in turn should presumably generate chiral edge currents and the condensate pairs also carry a relative orbital angular momentum (OAM) $l_z = M\hbar$ with reduced Planck constant $\hbar$ \cite{Anderson:1961,Volovik:1975,Leggett:1975,Ishikawa:1977,Cross:1977,Leggett:1978,McClure:1979,Sigrist:1990,Sauls:1994,Kita:1998,Matsumoto:1999,Matsumoto:1999_errata,Furusaki:2001,Stone:2004,Stone:2008,Sauls:2011,Byun:2018}. However, while chiral superconductors more generally belong to the class of integer quantum Hall systems \cite{Volovik:1988,Volovik:1992,Read:2000}, the chiral currents (and OAM) are actually not topologically protected \cite{Volovik:1988,Black-Schaffer:2012}, unlike in e.g.~Chern insulators. Specifically, charge is not a conserved quantity in superconductors, making all charge-related quantities unprotected, and chiral superconductors also lack a proper Chern-Simons action related to a current-current correlation but instead correspond to a current-density correlation \cite{Furusaki:2001,Stone:2004}. As a result, the overall contribution of the edge modes to the OAM and chiral current, i.e.~both charge and mass currents, have been shown to be highly variable and non-intrinsic, depending on e.g.~boundary conditions, impurities, gap anisotropy, pairing symmetry, and band effects \cite{Ashby:2009,Sauls:2011,Bouhon:2014,Huang:2014,Huang:2015,Tada:2015,Volovik:2015:b,Ojanen:2016,Suzuki:2016,Suzuki:2017,Goryo:2017,Wang:2018,Tada:2018,Sugiyama:2020,Nie:2020,Suzuki:2022,Suzuki:2023}. Several of these studies have shown that both the OAM and the total current per edge (i.e.~integrated current density) are typically much smaller in chiral $d$-wave superconductors compared to in chiral $p$-wave superconductors. In fact, it was shown that with a sharp confining potential in 2D, the OAM and total current even vanish at low temperature in the BCS limit ($|\Delta| \ll \EF$ with Fermi energy $\EF$) for all chiral states, except the isotropic chiral $p$-wave state \cite{Huang:2014,Tada:2015}. A vanishing current has also been found in certain chiral $p$-wave systems \cite{Volovik:2015:b} and in the microscopic limit on e.g.~a square lattice \cite{Huang:2014}, although chiral $d$-wave superconductivity on the honeycomb lattice still generate finite edge charge currents \cite{Black-Schaffer:2012}. For more details, see Ref.~\cite{Nie:2020} and references therein. This severe or even complete suppression of the edge currents, hampers experimental verification and characterization of proposed chiral $d$-wave superconductors, since the chiral currents are typically taken as a fingerprint of chiral superconductivity. It therefore becomes important to identify situations where the currents can be generally enhanced, or to find alternative fingerprints \cite{Holmvall:2023:cv,Holmvall:2023:robust}. 
In this work we focus on the former scenario, showing that the chiral charge currents and orbital magnetic moment (OMM) (i.e.~the charged analogue of the OAM) can in fact be enhanced and even have a maximum at low temperatures in the BCS limit, due to mesoscopic finite-size effects. Moreover, with these being quantities related to electric charge and magnetism, they offer straightforward possibilities for experimental detection. 

In particular, in this work we study how the chiral charge currents and OMM depend on the system radius $\mathcal{R}$, temperature $T$, and spontaneous Meissner screening (self-screening) in mesoscopic chiral $d$-wave superconductors. We find two size regimes dominated by completely different physical behavior. For large system sizes, $\mathcal{R} > 20\xi_0$, we find a $1/\mathcal{R}$ scaling for the total charge current and OMM, thus reproducing previous results of vanishing current in semi-infinite systems (i.e.~as $\mathcal{R}\to\infty$) \cite{Nie:2020}. Here, our natural length unit is $\xi_0 \equiv \hbar\vF / 2\pi\kB\Tc$, which is the effective superconducting coherence length \footnote{The coherence length and penetration depth are generally not constants, but may depend on e.g.~temperature, Fermi velocity, energy, and can vary with spatial inhomogeneities. More appropriately, $\lambda_0$ and $\xi_0$ are our natural length scales.} over which superconducting phenomena typically vary, with Fermi velocity $\vF(\vpF) = |\vvF(\vpF)|$ on the FS and Boltzmann's constant $\kB$. However, in smaller systems $\mathcal{R} \leq 20\xi_0$, we find that the spectrum is highly modified by hybridization between opposite edges and current conservation enforcing a faster suppression of the current density as function of distance from the edge. 
The combination of these effects even produce a sign reversal of the total charge current below $\mathcal{R} < 10\xi_0$. The existence of these two distinct finite-size regimes lead to a strongly enhanced charge current and OMM with local maxima at $\mathcal{R} \approx 20\xi_0$ and $\mathcal{R} \approx 10\xi_0$, respectively, and also notably with a maximum at low temperatures, which is in stark contrast to the vanishing current previously reported at the lowest temperatures for larger systems \cite{Wang:2018}. We further find that extremely strong self-screening causes a suppression of the charge current and OMM, but also a strong enhancement of the local induced magnetic-flux density which can be measured with magnetometry techniques e.g.~via a superconducting quantum interference device (SQUID) \cite{Geim:1997,Bolle:1999,Tsuei:2000,Morelle:2004,Kirtley:2005,Khotkevych:2008,Bleszynski-Jayich:2009,Kokubo:2010,Bert:2011,Jang:2011,Vasyukov:2013,Curran:2014,Kirtley:2016,Ge:2017,BishopVanHorn:2019,Persky:2022}. Our results thus demonstrate generically finite charge currents and OMM in chiral $d$-wave superconductors enhanced by mesoscopic finite-size effects, specifically edge-edge hybridization and effects of current conservation. As a consequence, mesoscopic superconductors may be a viable route to experimental verification of chiral superconductivity.

The rest of this work is outlined as follows. We present our theoretical model and methods in Sec.~\ref{sec:model}. Section~\ref{sec:chiral_groundstate} gives a background with general properties of chiral $d$-wave superconductivity and its order parameter. Section~\ref{sec:chiral_ground_state:edge_modes} gives a general background of chiral edge modes and compares them with flat bands of Andreev bound states (ABS), followed by a study of edge-edge hybridization in finite systems. Section~\ref{sec:chiral_ground_state:currents} presents our main results, consisting of the system-size dependence and temperature dependence of the charge-current density, total charge current per edge, and OMM. Section~\ref{sec:chiral_ground_state:induction} studies the effects of self-screening on these quantities for different system sizes. The work is concluded in Sec.~\ref{sec:conclusions}. Appendices contain further details on our theoretical formalism, analytic calculations in bulk and at interfaces, and additional numerical results.

\section{Model and methods}
\label{sec:model}
Chiral $d$-wave superconductivity has been proposed in a number of materials, with widely different atomic structures and properties \cite{Black-Schaffer:2007,Black-Schaffer:2014:b,Can:2021:a,Can:2021:b,Venderbos:2018,Su:2018,Fidrysiak:2018,Xu:2018,Kennes:2018,Liu:2018,Gui:2018,Wu:2019,Fischer:2021,Biswas:2021,Biswas:2013,Fischer:2014,Ueki:2019,Ueki:2020,Kasahara:2007,Kasahara:2009,Shibauchi:2014,Iguchi:2021,Takada:2003,Kiesel:2013,Yamanaka:1998,Kuroki:2010,Ming:2023,Gong:2017,Hosseinabadi:2019}.
In this work, we aim to study properties that are naturally emergent from the chiral $d$-wave pairing symmetry. This section describes our model and theoretical framework.

\subsection{Model}
\label{sec:model:model}
We are primarily interested in the combined influence of mesoscopic finite-size effects, temperature, and spontaneous Meissner screening on the steady-state charge current and its signature in spin-singlet chiral $d$-wave superconductors. We therefore appropriately assume equilibrium and spin degeneracy without additional spin-orbit interactions, instead leaving such effects as an interesting outlook. Since many proposed chiral $d$-wave superconductors are layered materials \cite{Can:2021:a,Can:2021:b,Biswas:2013,Fischer:2014,Ueki:2019,Ueki:2020,Ming:2023,Venderbos:2018,Su:2018,Fidrysiak:2018,Xu:2018,Kennes:2018,Liu:2018,Gui:2018,Wu:2019,Fischer:2021,Gong:2017,Hosseinabadi:2019} and some of these are weakly coupled, we consider weak-coupling superconductivity in 2D aligned with the $xy$-plane in our coordinate system, further modeling a cylindrically symmetric FS such that there is translational invariance along $\vzhat$ \cite{Graf:1993}. For clarity, we further assume clean systems with specular surfaces (i.e.~perfect specular reflection boundary conditions that conserve the surface-parallel momentum \cite{SuperConga:2023}), but effects of diffuse scattering are also briefly discussed. We do not consider the possibility of boundary-enhanced superconductivity as studied in Refs.~\cite{Samoilenka:2020,Samoilenka:2021,Hainzl:2022,Roos:2023,Roos:2023:b} which is primarily relevant for temperatures $T$ close to the critical temperature $\Tc$, while the most interesting effects we find happen in the range $T \in [0,\Tc/2]$. We consider effects of spontaneous Meissner screening of the chiral currents, due to a finite London-penetration depth $\lambda_0 \equiv \sqrt{c^2/\left(4\pi e^2 \vF^2\NF\right)}$ setting the effective length-scale of magnetic screening and induction, where $c$ is the speed of light, $e=-|e|$ the elementary charge, and $\NF$ the normal-state density of states (per spin) at the Fermi level. We assume type-II superconductivity,
appropriate for layered materials, thin films, non-elemental materials or most unconventional superconductors. Type-II superconductivity is typically quantified via the Ginzburg-Landau parameter $\kappa \equiv \lambda_0/\xi_0 > 1/\sqrt{2}$ (or more appropriately via the critical fields)  \cite{Kogan:2014:a,Kogan:2014:b,Ooi:2021,Prozorov:2022}.

\subsection{Quasiclassical theory: separation of scales}
\label{sec:model:quasiclassics}
We focus our attention on the low-energy physics close to the FS, where superconductivity is most important, and note that there is typically a separation in the relevant length- and energy-scales in many superconductors. In particular, the superconducting gap $|\Delta|$ is usually much smaller than the Fermi energy $\EF$, while the superconducting coherence length $\xi_0$ is often much larger than both the Fermi wavelength $\hbar/\pF$ and the atomic length scale $a_0$. In such superconductors, the low-energy (long-wavelength) physics can be separated from the high-energy (short-wavelength) contributions. The quasiclassical theory of superconductivity \cite{Eilenberger:1968,Larkin:1969,Serene:1983,Shelankov:1985,Nagato:1993,Eschrig:1994,Schopohl:1995,Schopohl:1998,Eschrig:1999,Grein:2013,SuperConga:2023} is a controlled expansion in the resulting small parameters, e.g.~$|\Delta|/\EF$ and $\hbar/(\pF\xi_0)$, and where the leading-order terms are obtained from the low-energy bands close to the FS. Higher-energy contributions can still be inserted into the theory, e.g.~through microscopic boundary conditions \cite{Zaitsev:1984,Kieselmann:1987,Millis:1988,Eschrig:2000,Shelankov:2000,Fogelstrom:2000,Zhao:2004,Eschrig:2009}.

In quasiclassical theory, spatial variations typically occur on the mesoscopic length scales set by $\xi_0$ and $\lambda_0$. In particular, the propagation of quasiparticles and superconducting pairs are captured by the quasiclassical propagators $g(\vpF, \vR; z)$ and $f(\vpF, \vR; z)$, respectively. Here, $z$ is the complex-valued energy of the corresponding propagator. In this work, the retarded propagator $g^{\mathrm{R}}(\vpF,\vR;\varepsilon)$ with $z^{\mathrm{R}} \equiv \varepsilon + i\delta$ (real energy $\varepsilon$ and small positive broadening $\delta$) is used to calculate the local density of states (LDOS)
\begin{align}
    \label{eq:model:ldos}
    N(\vR; \varepsilon) = -\frac{2\NF}{\pi} \left\langle \operatorname{Im}\left[ g^{\mathrm{R}}(\vpF, \vR; \varepsilon)\right] \right\rangle_{\vpF},
\end{align}
with total density of states (DOS) $N(\varepsilon) = \int_{\mathcal{A}} d^2\vR N(\vR,\varepsilon)$ integrated over the system area $\mathcal{A}$, and where the angle brackets denote the FS average \cite{Graf:1993}
\begin{align}
    \label{eq:model:fs_average}
    \left\langle \dots \right\rangle_{\vpF} = \frac{1}{\NF} \oint_\FS \frac{\mathrm{d} p_\mathrm{F}}{(2\pi\hbar)^2|\vvF(\vpF)|} (\dots),
\end{align}
which is a line integral in 2D averaging over the Fermi-momentum direction. Similarly, the normal-state density of states (per spin) can further be expressed as
\begin{align}
    \label{eq:model:normal_dos}
    \NF = \oint_\FS \frac{\mathrm{d} \pF}{(2\pi\hbar)^2|\vvF(\vpF)|},
\end{align}
and the Fermi velocity is parametrized via the FS according to $\vvF(\vpF) = \nabla_{\vp}\varepsilon(\vp)|_{\vp=\vpF}$, with dispersion $\varepsilon(\vp)$.
Matsubara propagators $g^{\mathrm{M}}(\vpF,\vR;\varepsilon_n)$ and $f^{\mathrm{M}}(\vpF,\vR;\varepsilon_n)$ are used for all other quantities as described in the following, evaluated on the imaginary axis via the Matsubara energies $z^{\mathrm{M}} \equiv i\varepsilon_n = i\pi\kB T(2n+1)$ with temperature $T$ and integer $n$. In Nambu (particle-hole) space, we construct
\begin{align}
\label{eq:model:green_function}
\hat{g}(\vpF, \vR; z) = 
\begin{pmatrix}
    g(\vpF, \vR; z) & f(\vpF, \vR; z)\\
    -\tilde{f}(\vpF, \vR; z) & \tilde{g}(\vpF, \vR; z)
\end{pmatrix},
\end{align}
where ``tilde'' denotes particle-hole conjugation
$\tilde{\alpha}(\vpF, \vR; z) = \alpha^*(-\vpF, \vR; -z^*)$. We briefly drop the arguments $(\vpF, \vR; z)$ for a more compact notation and follow the Eilenberger formulation where $\hat{g}$ is solved from the Eilenberger equation \cite{Eilenberger:1968}
\begin{align}
    \label{eq:model:eilenberger}
    i\hbar\vvF\cdot\boldsymbol{\nabla}\hat{g} + \left[z\hat{\tau}_3 - \hat{h},\hat{g}\right] = 0,
\end{align}
with the normalization condition $\hat{g}^2 = -\pi^2\hat{\tau}_0$.
Here, $\hat{\tau}_i$ are the $4\times4$ Pauli-spin matrices in Nambu space with identity matrix $\hat{\tau}_0$, while $\hat{h}$ are the self-energies, which we divide into diagonal ($\hat{\Sigma}$) and off-diagonal ($\hat{\Delta}$) parts,
\begin{align}
    \label{eq:model:self_energy}
    \hat{h} = \hat{\Sigma} + \hat{\Delta} = \begin{pmatrix}
    \Sigma & \Delta \\
    \tilde{\Delta} & \tilde{\Sigma}
    \end{pmatrix}.
\end{align}
Here, $\Delta$ is the mean-field superconducting order parameter as described in Sec.~\ref{sec:model:superconductivity}, while $\Sigma$ generally includes different interactions and other diagonal self-energies. In this work, $\Sigma$ captures the electrodynamics, see Sec.~\ref{sec:model:electrodynamics}.

\subsection{Superconductivity}
\label{sec:model:superconductivity}
Assuming an even-parity spin-singlet superconductor, the effective pairing interaction $V(\vpF,\vpFp)$ is decomposed into the symmetry channels of the corresponding crystallographic point group \cite{Yip:1993},
\begin{align}
    \label{eq:model:pairing_interaction}
    V(\vpF,\vpFp)=\sum_{\Gamma} V_\Gamma \eta_\Gamma(\vpF)\eta^\dagger_\Gamma(\vpFp).
\end{align}
Here, $\Gamma$ labels the even-parity spin-singlet irreducible representations, $V_\Gamma$ the pairing strength of the respective symmetry channel, and $\eta_\Gamma(\vpF)$ is the basis function encoding the pairing symmetry on the FS. Similarly, the total superconducting order parameter $\Delta(\vpF,\vR)$ is written as
\begin{align}
    \label{eq:model:order_parmeter:irreducible_representation}
    \Delta(\vpF,\vR) = \sum_\Gamma \Delta_\Gamma(\vpF,\vR) = \sum_\Gamma|\Delta_\Gamma(\vR)|e^{i\chi_\Gamma(\vR)}\eta_\Gamma(\vpF)
\end{align}
where each symmetry channel is associated with an order parameter component $\Delta_\Gamma(\vpF,\vR)$ with amplitude $|\Delta_\Gamma(\vR)|$ and phase $\chi_\Gamma(\vR)$.
We use the efficient ``Ozaki summation'' \cite{Ozaki:2007} based on the Matsubara technique \cite{Matsubara:1955,Bruus:2004,Rammer:2007,Kopnin:2009,Mahan:2013} to solve the order parameter self-consistently from the superconducting gap equation
\begin{align}
    \label{eq:model:gap_equation}
    \Delta(\vpF,\vR) = \NF \kB  T\sum_n^{|\varepsilon_n|<\Omegac} \big\langle
                  V(\vpF,\vpFp)\,f(\vpFp,\vR;\varepsilon_n)
                   \big\rangle_{\vpFp},
\end{align}
where the cutoff $\Omegac$ is effectively the bandwidth of the pairing interaction \cite{Grein:2013}.

In this work, we study chiral $d$-wave superconductivity, requiring finite pairing channels $\Gamma \in \{d_{x^2-y^2}, d_{xy}\}$ with $\eta_{d_{x^2-y^2}}(\vpF) = \sqrt{2}\cos\left(2\thetaF\right)$ and $\eta_{d_{xy}}(\vpF) = \sqrt{2}\sin\left(2\thetaF\right)$,
where additional pair correlations of other symmetries are automatically included in the theory (e.g. $s$-wave), while the influence of other attractive pairing channels \cite{Black-Schaffer:2013} is left as an outlook. We assume a degenerate pairing with equal $V_\Gamma$ in both $d$-wave channels which is enforced by symmetry in any material with a three- or six-fold rotationally symmetric lattice \cite{Black-Schaffer:2014:b}, thus relevant for many of the recently proposed chiral $d$-wave superconductors \cite{Venderbos:2018,Su:2018,Fidrysiak:2018,Xu:2018,Kennes:2018,Liu:2018,Gui:2018,Wu:2019,Fischer:2021,Ming:2023,Biswas:2013,Fischer:2014,Ueki:2019,Ueki:2020}.
We thus consider a total order parameter of the form
\begin{align}
    \label{eq:model:order_parameter:dwaves}
    \Delta(\vpF,\vR) & = \Delta_{d_{x^2-y^2}}(\vpF,\vR) + \Delta_{d_{xy}}(\vpF,\vR).
\end{align}
To better highlight the chiral properties, it is useful to rewrite the superconducting order parameter from Eq.~(\ref{eq:model:order_parameter:dwaves}) in the eigenbasis $\eta_\pm(\vpF)$ of the orbital (orb) angular-momentum operator, $\Lzorb \equiv (\hbar/i)\partial_{\thetaF}$,
via the linear combination
\begin{align}
    \label{eq:chiral:op:plus_minus}
    \Delta(\vpF,\vR) & = \Dplus(\vpF,\vR) + \Dminus(\vpF,\vR),\\
    \label{eq:chiral:op:components:plus_minus}
    \Dpm(\vpF,\vR) & \equiv |\Dpm(\vR)|e^{i\chi_{\pm}(\vR)}\eta_{\pm}(\vpF),\\
    \label{eq:chiral:eta:plus_minus}
    \eta_{\pm}(\vpF) & \equiv \frac{\etaxtyt(\vpF) \pm i\etaxy(\vpF)}{\sqrt{2}} = e^{\pm i 2\thetaF}.
\end{align}
We see that $\Dpm$ corresponds to a relative phase shift $\pm\pi/2$ between the components $\Dxtyt$ and $\Dxy$. There is no loss in generality in this basis change (it is valid even in the non-chiral state, see Sec.~\ref{sec:chiral_groundstate} for definition of chiral state). Equating Eq.~(\ref{eq:model:order_parameter:dwaves}) with Eq.~(\ref{eq:chiral:op:plus_minus}) yields the transformation between the two parametrizations
\begin{align}
    |\Dpm(\vR)|e^{i\chi_\pm(\vR)} = \frac{1}{\sqrt{2}}\Big(|\Dxtyt(\vR)|e^{i\chixtyt(\vR)} \nonumber\\
    \label{eq:chiral:transform:delta_plusminus}
    \mp i|\Dxy(\vR)|e^{i\chixy(\vR)}\Big).
\end{align}

For comparison, we can generalize Eq.~(\ref{eq:chiral:op:components:plus_minus}) for a superfluid of angular order $M$ (often referred to as the chiral order), i.e. the Chern number \cite{Volovik:1997,Schnyder:2008,Hasan:2010,Qi:2011,Tanaka:2012,Graf:2013,Black-Schaffer:2014:b}, where in 2D even and odd $M$ correspond to spin-singlet and spin-triplet (e.g. $M=\pm1,\pm2,\pm3$ for $p,d,f$-wave respectively), $\Dpm(\vpF,\vR) = |\Dpm(\vR)|e^{i\chi_\pm(\vR)}e^{iM \thetaF}$.
Indeed, this state is an eigenstate of $\Lzorb \Dpm(\vpF,\vR) = l_{\textrm{orb}} \Dpm(\vpF,\vR)$ with eigenvalue $l_{\textrm{orb}} = \hbar M$
This corresponds to a condensate which carries an OAM along the $\pm \hat{z}$-axis \cite{Sauls:1994}. In contrast, a non-chiral state, like a nodal or nematic $d$-wave state \cite{Lothman:2022}, is not an eigenstate of $\Lzorb$. 
We from here on refer to $\Dxtyt$ and $\Dxy$ as the nodal components, and $\Dpm$ as the chiral components.

Finally, note that apart from applying a start guess to the order parameter, we do not constrain the order parameter components $\Delta_\Gamma$ or their phases $\chi_\Gamma$ in any way. Instead, we let them evolve completely independently in the self-consistency. This in principle allows the system to find a non-chiral state, but we always find the chiral state to be stable.

\subsection{Self-consistent electrodynamics}
\label{sec:model:electrodynamics}
Electrodynamical interactions coupling to the orbital degrees of freedom and caused by external magnetic fields (ext) or magnetic induction (ind) enter the diagonal self-energies $\hat{\Sigma}$ via
\begin{align}
    \label{eq:model:self_energy:diagonal}
    \hat{\Sigma} = -\frac{e}{c}\vvF(\vpF)\cdot\vA(\vR)\hat{\tau}_3,
\end{align}
where $\vA(\vR) = \vAext(\vR) + \vAind(\vR)$ is the electromagnetic gauge field.
The magnetic flux-density, $\vB(\vR) = \vBext(\vR) + \vBind(\vR)$,
is related to the vector potential via $\vB(\vR) = \vnabla \times \vA(\vR)$.
We do not consider external flux in this work, such that $\vBext = 0$ and $\vAext = 0$.
Still, $\vAind$ is an induced vector potential due to spontaneous charge currents in the system, related via Amp{\`e}re's law
\begin{align}
    \label{eq:model:ampere}
    \vnabla\times\vBind(\vR) = \vnabla \times \vnabla \times \vAind(\vR) = \frac{4\pi}{c}\vj(\vR).
\end{align}
The charge-current density is in turn given by
\begin{align}
    \label{eq:model:current_density}
    \vj(\vR) = 2e \NF \kB T\sum_{n}^{|\varepsilon_n|<\Omegac} \left\langle \vvF(\vpF)\, g(\vpF,\vR;\varepsilon_n) \right\rangle_{\vpF},
\end{align}
which we express in units of $j_0 \equiv \hbar|e|\vF^2\NF/\xi_0 = 2\pi|e|\kB\Tc\NF\vF$. Here $\vj(\vR)$ is the total charge-current density 
with contributions from both quasiparticles 
and superconducting pairs 
(hence $\vj$ is conserved with $\vnabla \cdot \vj = 0$).
The induction gives rise to an additional self-consistency equation, Eq.~(\ref{eq:model:ampere}), to be solved together with the self-consistent superconducting gap equation, Eq.~(\ref{eq:model:gap_equation}). We note that the induction is effectively weighted by a factor $(\lambda_0/\xi_0)^{-2} \ll 1$ \cite{Holmvall:2017}, which for extreme type-II superconductors ($\lambda_0 \gg \xi_0$) often makes the induction negligible when $\lambda_0 \gg \mathcal{R}$. Still, we always solve both $\Delta$ and $\vA$ fully self-consistently when $\lambda_0$ is finite, using Eqs.~(\ref{eq:model:gap_equation}) and (\ref{eq:model:ampere}). 

The orbital magnetic moment (OMM) $\vm = m_z\vzhat$ (per 2D layer) is computed from the charge-current density via \cite{SuperConga:2023}
\begin{align}
    \label{eq:model:magnetic_moment}
    \frac{\vm}{m_0} \equiv \frac{1}{2}\int_{\mathcal{A}} \frac{d^2\vR}{\mathcal{A}} \frac{\vR}{\xi_0} \times \frac{\vj(\vR)}{j_0},
\end{align}
with natural units $m_0 = N\hbar\frac{|e|}{m^*} = 2\muB N$
with Bohr magneton $\muB = \hbar|e|/2m^*$, particle number $N$, and effective quasiparticle mass $m^*$ defined via $\vpF = m^*\vvF$. We note the analogue between the OMM in Eq.~(\ref{eq:model:magnetic_moment}) and the orbital angular momentum (OAM) $\vL^{\mathrm{orb}} = \int_{\mathcal{A}} d^2\vR\, \vR \times \vjm(\vR)$, typically expressed in units $L_0=N\hbar/2$, with mass-current density $\vjm$ \cite{Sauls:2011}.

Finally, we introduce the area-averaged induced flux density (or equivalently total induced flux)
\begin{align}
    \label{eq:spontaneous_magnetization}
    \Phiind = \int_{\mathcal{A}} d^2\vR \Bind(\vR),
\end{align}
with flux quantum $\Phi_0 \equiv hc/2|e|$ and Planck constant $h$.

\subsection{Calculations details}
\label{sec:model:numerics}
We use the Riccati formalism \cite{Shelankov:1985,Nagato:1993,Schopohl:1995,Schopohl:1998,Eschrig:2000,Eschrig:2009,Grein:2013} described in Appendix~\ref{app:riccati} to solve the Eilenberger equation, Eq.~(\ref{eq:model:eilenberger}), numerically, specifically using the open-source framework SuperConga \cite{SuperConga:2023} which is free to download from its Gitlab repository \cite{SuperConga:repository} with extensive documentation \cite{SuperConga:documentation}. See Ref.~\cite{SuperConga:2023} for extensive implementation details. The Eqs.~(\ref{eq:model:eilenberger}), (\ref{eq:model:gap_equation}) and (\ref{eq:model:ampere}) are solved self-consistently in an iterative process by starting from an appropriate start guess, and proceeding until the global error $\epsilon_{\mathrm{G}}$ of quantity $O_i$ at iteration number $i$ is $\epsilon_{\mathrm{G}} = \left \| O_i-O_{i-1} \right \|_2/\left \|O_{i-1} \right \|_2 < \epsilon_{\mathrm{tol}}$
for $O \in \{\Delta(\vpF,\vR),\vA(\vR),\vj(\vR),\Omega\}$ and tolerance $\epsilon_{\mathrm{tol}}$. In the present work, we set the tolerance to $\epsilon_{\mathrm{tol}} = 10^{-7}$, use a spatial resolution of $20$ discrete points per coherence length, and use an energy cutoff $\Omegac \gtrsim 100 \kB\Tc$. We parametrize the FS with $256$ discrete points, except at low temperature or when computing the LDOS, in which cases we use $512$ and $720$ discrete points, respectively. We find these values sufficient, as we do not notice any difference with finer resolution.

In order to investigate the influence of finite geometry, we simulate discs with radii $\mathcal{R} \in [5,200]\xi_0$, temperatures $T\in[0.01,0.99]\Tc$, and penetration depths $\lambda_0$ via $\kappa \equiv \lambda_0/\xi_0 \in [1,\infty)$. We keep the same spatial resolution when changing $\mathcal{R}$, and keep all parameters fixed during the process of converging towards the self-consistent solution.
In order to avoid divergences in the LDOS calculation, we use a small, phenomenological, energy broadening $\delta \sim 0.03\kB\Tc$ in all the LDOS calculations.

In order to improve the understanding and interpretation of our numerical results, we provide supplementary analytic calculations in Appendices~\ref{app:analytics:bulk} and \ref{app:analytics:surface} for bulk and surface scenarios, respectively. 

\section{Background: chiral $d$-wave superconductivity}
\label{sec:chiral_groundstate}
To set the stage for our studies of the chiral edge modes and charge currents in the subsequent sections, we here review the basic properties of a chiral $d$-wave order parameter in a finite system.

A chiral $d$-wave superconductor is characterized by the existence of two degenerate ground states of opposite chiralities, here denoted $\Dplus(\vpF,\vR)$ and $\Dminus(\vpF,\vR)$ defined in Eqs.~(\ref{eq:chiral:op:plus_minus})--(\ref{eq:chiral:eta:plus_minus}). At the onset of chiral superconductivity, the bulk superconductor spontaneously chooses one of the two chiral states ($\Dpm$) as the dominant order, while the other state ($\Dmp$) becomes subdominant and vanishes asymptotically \cite{Hess_Tokuyasu_Sauls:1989}. Hence, the order parameter takes the approximate form $\Delta(\vpF,\vR) \approx \Dpm(\vpF,\vR)$ in these domains, which leads to a fully gapped state \cite{Kallin:2012,Kallin:2016,Mizushima:2016,Volovik:2019}. Since the two states $\Dplus$ and $\Dminus$ are related via time reversal symmetry, the appearance of a dominant chirality leads to spontaneous time-reversal symmetry-breaking 
\cite{Rice:1995,Volovik:1997}.
The overall global features of the superconductor is determined by this dominant component. The subdominant component can still carry additional information about local features as it is induced within a few coherence lengths close to spatial inhomogeneities, such as interfaces, defects and vortices \cite{Heeb:1990,Ichioka:2002}. Hence, in the presence of such inhomogeneities, the order parameter is not in a pure chiral state, and takes the more general form $\Delta(\vpF,\vR) = \Dplus(\vpF,\vR) + \Dminus(\vpF,\vR)$ from Eq.~(\ref{eq:chiral:op:plus_minus}).
\begin{figure}[t!]
	\includegraphics[width=\columnwidth]{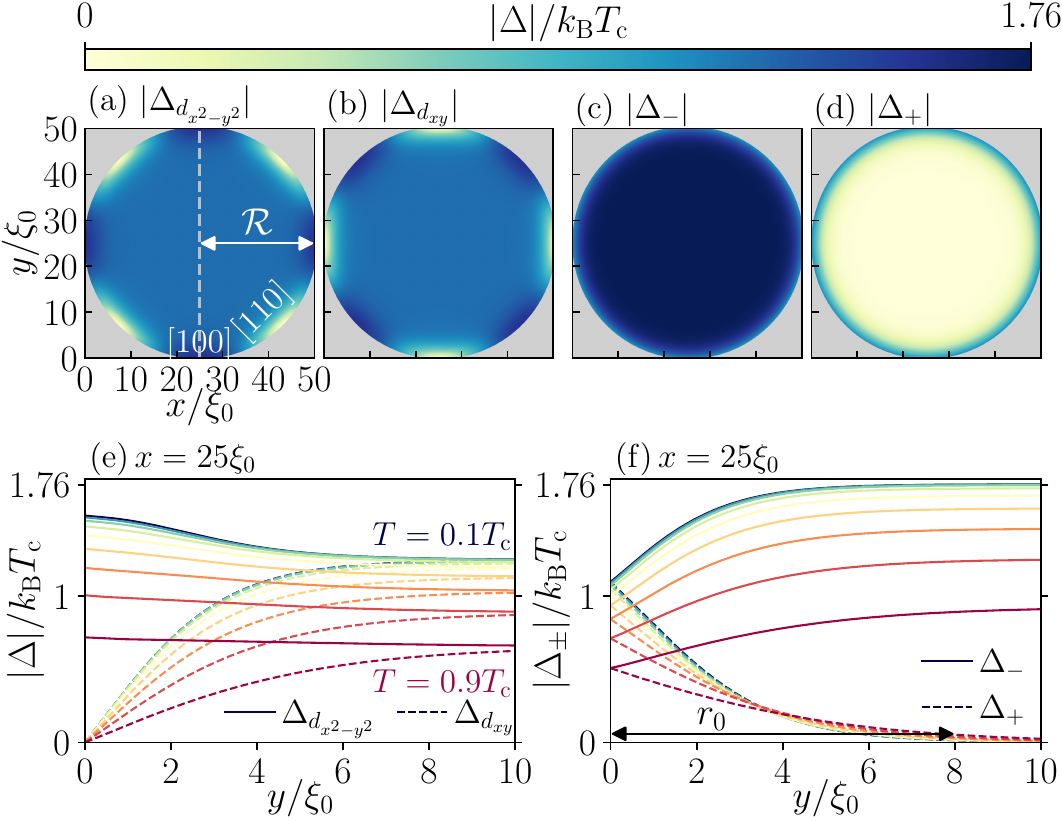}
	\caption{ Ground-state order parameter of a chiral $d$-wave superconductor with disc radius $\mathcal{R}=25\xi_0$, dominant chirality $\Dminus$, $T=0.1\Tc$, $\lambda_0=\infty$.
        (a,b) Amplitudes of nodal components, and (c,d) chiral components.
	(e,f) Temperature dependence of the order parameter amplitudes, as a function of the distance from a $[100]$ edge, along the dashed line ($x=25\xi_0$) in (a). Temperatures vary from $T=0.1{\Tc}$ (blue) to $T=0.9{\Tc}$ (red), in steps of $\Delta T = 0.1\Tc$.
	}
	\label{fig:chiral_ground_state:op}
\end{figure}
In Fig.~\ref{fig:chiral_ground_state:op}, we demonstrate the behavior of the chiral $d$-wave order parameter solved self-consistently as described in Sec.~\ref{sec:model}, in a mesoscopic system shaped like a disc with dominant component $\Dminus$, but where the subdominant component $\Dplus$ arises close to the edge. For comparison, we show results with both parametrizations (a,b) $\Delta = \Dxtyt + \Dxy$ and (c,d) $\Delta = \Dplus + \Dminus$, while (e,f) show the spatial dependence  of the order parameter amplitudes at different temperatures, starting at the edge ($y=0$) where we find $|\Dplus| = |\Dminus|$. This degeneracy is related to the surface suppression of one of the nodal components ($\Dxy$ at a $[100]$ interface) due to its sign change, leading to surface bound states that locally enhances the other nodal component ($\Dxtyt$) \cite{Sauls:2011}. We study edge states in detail in Sec.~\ref{sec:chiral_ground_state:edge_modes}.

We note that bulk chiral superconductivity recovers over a characteristic and effective length scale $\decayLength \approx {6\text{--}10\xi_0}$ (depending on e.g.~temperature),
typical for many inhomogeneous superconducting phenomena in clean and mesoscopic systems \cite{Lofwander:2001,Holmvall:2017,Holmvall:thesis:2019}. We trace the length scale $\decayLength$ as naturally emerging from the typical spatial dependence for the ballistic propagators in non-uniform environments. For example, at a $[100]$ interface with translational invariance along $x$ (or equivalently large radius of curvature compared to $\xi_0$), this spatial dependence is captured by the appearance of factors $\exp(-y/\xiy)$ in the quasiparticle propagators and pair propagators. Here, $\xiy$ denotes the effective ``non-constant coherence length'' \cite{Holmvall:thesis:2019}, which we derive analytically in Appendix~\ref{eq:app:riccati:inhomogeneous:gamma} as
\begin{equation}
    \label{eq:xiy:main}
    \xiy(\vpF;\varepsilon_n;T) = \frac{\hbar\left|\vFy(\vpF)\right|}{2\sqrt{|\Delta(\vpF;T)|^2 + \varepsilon_n^2(T)}},
\end{equation}
where $\vFy(\vpF)$ is the $y$-component (i.e.~perpendicular to the surface) of the Fermi velocity $\vvF(\vpF)$. To compute most quantities, the propagators are averaged over the FS (as described in Sec.~\ref{sec:model}), which means that the factor $\exp(-y/\xiy)$ essentially capture multiple confinement scales $\xiy(\vpF;\varepsilon_n;T)$ \cite{Sauls:2011}, and illustrates how the boundaries couple the real-space ballistic trajectories to the momentum-space dependence on the FS, as well as the temperature- and energy-dependence. In averaging the propagators over the FS and summing over Matsubara energies, the temperature-dependent decay length $\decayLength$ over which bulk superconductivity recovers emerges through the exponential factors. However, due to the non-trivial averages and sums, a direct relation between $r_0$ and $\xiy$ seems intractable, and we therefore treat $r_0$ as an emergent length scale which ultimately depends on the superconducting coherence length via $\xiy$.
Finally, we note that as the system radius $\mathcal{R}$ becomes comparable to $\decayLength$, there is a strong wave function overlap between opposite pairbreaking surfaces, consequently causing significant reduction of both nodal $d$-wave components in the entire system, see Appendix~\ref{app:additional_results:OP} where we include additional results explicitly showing how the order parameter changes with $\mathcal{R}$. In Sec.~\ref{sec:chiral_ground_state:edge_modes} we show that this overlap is also associated with an edge-edge hybridization of the spectrum and the chiral edge modes. Thus, these are mesoscopic finite-size effects that strongly influence the physics in mesoscopically sized superconductors, as we illustrate throughout this work.

\section{Chiral edge modes}
\label{sec:chiral_ground_state:edge_modes}
Chiral superfluids are known to host chiral edge modes (Weyl fermions \cite{Volovik:1992,Volovik:2020}) which are related to the bulk Chern number via the bulk-boundary correspondence \cite{Volovik:2003,Mizushima:2016,Volovik:2019}. To provide further background for the mesoscopic finite-size effects on the chiral currents, we begin this section by revisiting the well-studied problem of the chiral spectrum and explore the edge modes in a chiral $d$-wave superconductor. Specifically, we focus on the temperature dependence of the LDOS and point out clear differences against surface-bound states in a regular $d$-wave superconductor. Finally, we compute the LDOS for different system sizes and illustrate how the edge modes are influenced by mesoscopic finite-size effects, specifically edge-edge hybridization.

\subsection{Edge modes and bulk LDOS}
\label{sec:chiral_ground_state:edge_modes:spectrum}
\begin{figure}[t!]
	\includegraphics[width=\columnwidth]{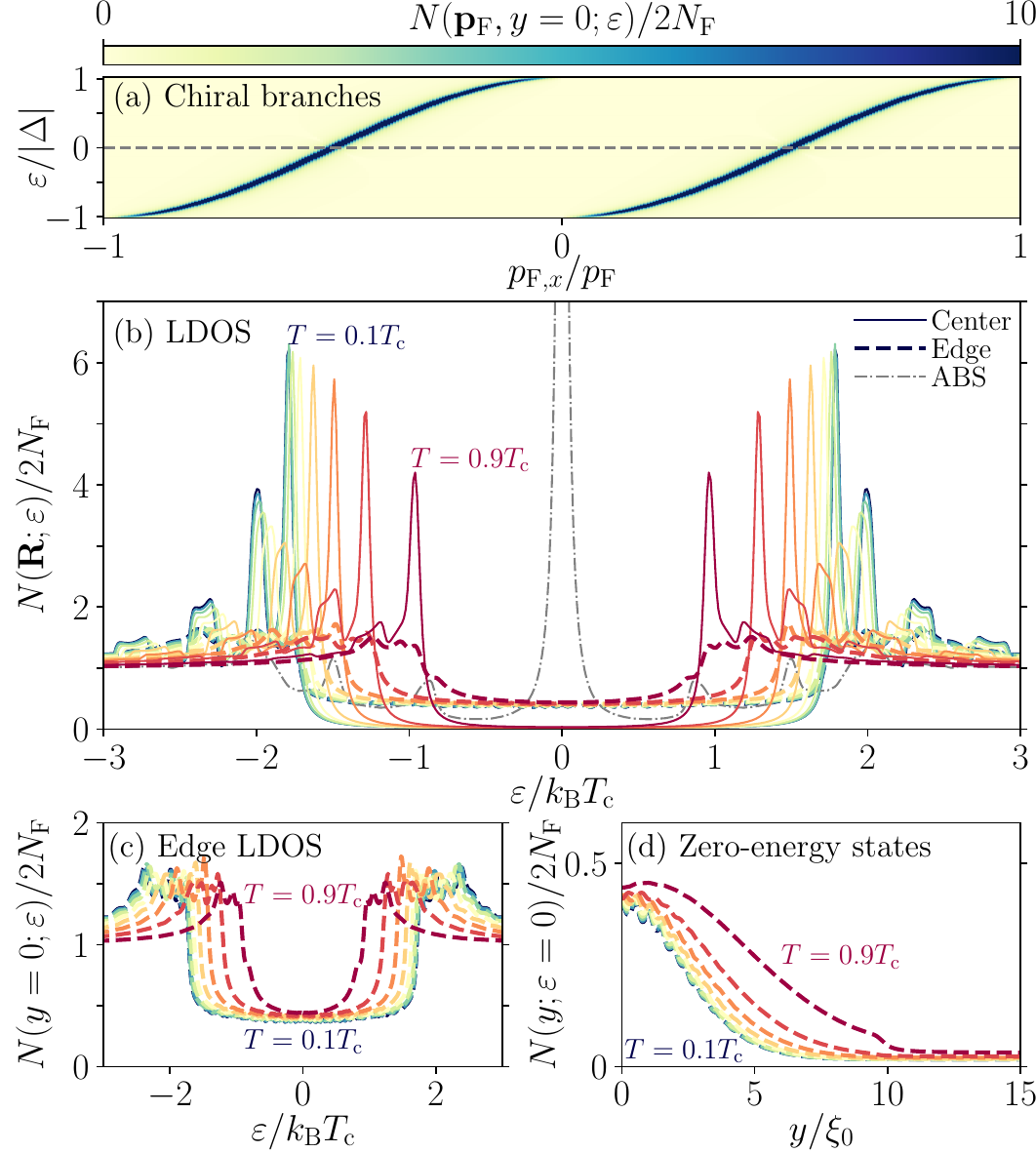}
	\caption{(a) Subgap spectrum ${N(\vpF;\varepsilon<|\Delta|)}$ at the edge (${y=0}$) of a semi-infinite chiral $d$-wave superconductor, with surface-parallel momentum $p_{\parallel} = \pFx = \pF\cos(\thetaF)$.
	(b) Self-consistent LDOS $N(\vR;\varepsilon)$ at the center (solid) and edge (dashed) of a disc with radius $\mathcal{R}=30\xi_0$, $\lambda_0=\infty$. Colors denote temperature from $T=0.1{\Tc}$ (blue) to $T=0.9{\Tc}$ (red). In contrast, dash-dotted line are the flat-band ABS at a $[110]$ interface in a nodal $d_{x^2-y^2}$-wave superconductor. (c) Zoom of the edge LDOS in (b). (d) Spatial dependence of the chiral edge modes at zero energy.}
	\label{fig:chiral_spectrum}
\end{figure}
\begin{figure*}[t!]
	\includegraphics[width=\textwidth]{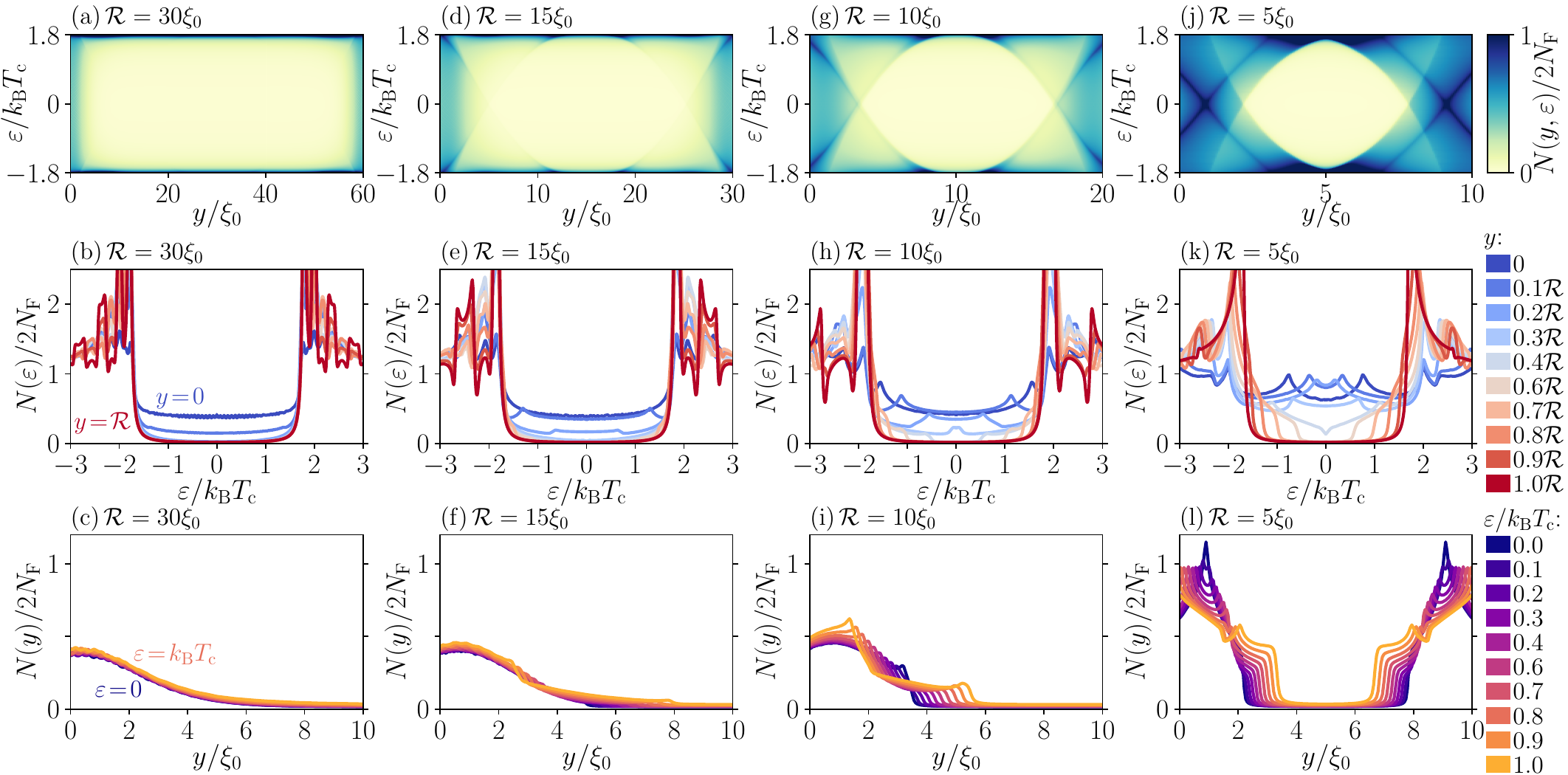}
	\caption{First row: subgap LDOS versus energy $\varepsilon$ and coordinate $y$ across the disc diameter at $T=0.1\Tc$, $\lambda_0=\infty$, and different disc radii $\mathcal{R}$ (columns), with bulk gap $\Delta_0\approx1.76\kB\Tc$. Second row: same but line cuts at fixed $y$ from the disc edge $y=0$ (blue) to the disc center $y=\mathcal{R}$ (red). Third row: same but line cuts at fixed $\varepsilon$ from $\varepsilon=0$ (purple) to $\varepsilon=\kB\Tc$ (orange).}
	\label{fig:radius_chiral_edge_modes}
\end{figure*}
A chiral superfluid with Chern number $M$ has $|M|$ topologically protected edge modes \cite{Volovik:1997,Schnyder:2008,Hasan:2010,Qi:2011,Tanaka:2012,Graf:2013,Black-Schaffer:2014:b}. In Fig.~\ref{fig:chiral_spectrum}(a) we show the subgap spectrum at a semi-infinite surface of a chiral $d$-wave superconductor (derived analytically without self-consistency in Appendix~\ref{app:analytics:surface:specular}, while all subsequent results in the main text are numeric and fully self-consistent). There are $|M|=2$ chiral branches with dispersion $\varepsilon(p_\parallel)$ as a function of the surface-parallel momentum $p_\parallel$ \cite{Black-Schaffer:2012}. The broken time-reversal symmetry of the chiral state is explicitly seen, since for each occupied state $\varepsilon(p_\parallel)$, there is no time-reversed partner at $\varepsilon(-p_\parallel)$ \cite{Rice:1995,Volovik:1997}.
Figure~\ref{fig:chiral_spectrum}(b) shows the fully self-consistent LDOS at the center ${(x,y)=(\mathcal{R},\mathcal{R})}$ and edge ${(x,y)=(\mathcal{R},0)}$ in a disc with radius $\mathcal{R}=30\xi_0$, at different temperatures. In the center, a bulk-like state $\Delta(\vpF,\vR) \approx \Dminus(\vpF,\vR)$ is established. The bulk state is fully gapped, see also Appendix~\ref{app:analytics:bulk} for analytic derivation of the bulk gap. At the edge, the chiral edge modes generate a nearly constant subgap LDOS, see also the zoom of the edge LDOS in Fig.~\ref{fig:chiral_spectrum}(c). Finally in Fig.~\ref{fig:chiral_spectrum}(d) we show that at zero energy, the edge modes decay into the bulk over the same characteristic distance $\decayLength \sim 6\text{--}10\xi_0$ as the order parameter variations in Sec.~\ref{sec:chiral_groundstate}. Note that at higher temperatures, the superconducting gap is suppressed due to thermal excitations, both at the edge and in the bulk. This leads to a larger effective coherence length, Eq.~(\ref{eq:xiy:main}), and the edge modes consequently reach further into the bulk at higher temperatures.

\subsection{Comparison with nodal $d$-wave state}
\label{sec:chiral_ground_state:edge_modes:nodal_dwave}
In contrast to the full gap of the bulk chiral $d$-wave state, a regular $d$-wave superconductor has a gapless nodal spectrum and can host surface ABS \cite{Hu:1994,Jian:1994,Lofwander:2001,Vorontsov:2018}, the latter we also include in Fig.~\ref{fig:chiral_spectrum}(b) for comparison (gray dash-dotted line). There exist several important differences between the ABS in a nodal $d$-wave superconductor, and the chiral edge modes in a chiral $d$-wave superconductor. First, the ABS correspond to a flat band (in $k$-space) \cite{Sato:2011}, while the chiral edge modes are dispersive \cite{Black-Schaffer:2012}.
Second, the ABS peak typically appears as a Lorentzian function centered around the Fermi level and with a width influenced by e.g. boundary conditions and typical energy-broadening effects \cite{Lofwander:2001,Seja:2021,Seja:2022:thermopower,Seja:2022:current_injection}, while the chiral edge mode has a nearly constant subgap LDOS.
Third, we find that the zero-energy LDOS of the ABS is roughly two orders of magnitude larger than the chiral edge mode LDOS in Fig.~\ref{fig:chiral_spectrum}(b).
Fourth, the chiral LDOS is largely independent of the surface orientation, while the ABS in a $d_{x^2-y^2}$-wave superconductor appear mainly at the pairbreaking $[110]$-interfaces but are absent at $[100]$ interfaces \cite{Lofwander:2001}.
Fifth, the ABS are highly degenerate and thermodynamically unstable towards spontaneous symmetry breaking \cite{Matsumoto:1995,Fogelstrom:1997,Sigrist:1998,Barash:2000,Lofwander:2000,Honerkamp:2000,Vorontsov:2009,Potter:2014,Nagai:2017,Nagai:2018}, possibly triggering a phase transition into a state known as a ``phase crystal'' \cite{Holmvall:2020} at a relatively high temperature $T^*\sim0.2\textrm{--}0.5\Tc$ \cite{Hakansson:2015,Holmvall:2018b,Holmvall:2019,Holmvall:2020,Wennerdal:2020,Chakraborty:2022,Seja:2022:FEM,Bonetti:2023}. We find that the low spectral weight in the chiral $d$-wave system in Fig.~\ref{fig:chiral_spectrum}(b) does not facilitate such a thermodynamic instability. 
Finally, we point out that the ABS have a significant paramagnetic response \cite{Xu:1995,Higashitani:1997,Walter:1998,Suzuki:2014,Suzuki:2015,Suzuki:2023},
which we find can be very different from the response of the chiral edge modes (not shown).

\subsection{Mesoscopic finite-size effects: hybridization}
\label{sec:chiral_ground_state:edge_modes:mesoscopics}
Next, we investigate how the finite size influences the subgap LDOS, and in particular the chiral edge modes. Figure~\ref{fig:radius_chiral_edge_modes} shows the subgap LDOS versus energy and coordinate along the diameter (top row) and line cuts through the same data at fixed coordinates (middle row) and at fixed energies (bottom row). Different columns correspond to different disc radii $\mathcal{R}$.
Based on these results we find two different regimes with significantly different behavior: $\mathcal{R} > \regimeLengthXi$, and $\mathcal{R} < \regimeLengthXi$. 

For all radii $\mathcal{R} > \regimeLengthXi$, the edge modes behave the same as in Fig.~\ref{fig:radius_chiral_edge_modes}(a-c): the edge modes decay into the bulk over the same characteristic length scale $\decayLength \approx 6\text{--}10\xi_0$ as the order parameter in Figs.~\ref{fig:chiral_ground_state:op}(i-j). We note that this is similar to the spatial dependence of ABS at pairbreaking $[110]$ interfaces of a nodal $\dxtyt$-wave superconductor \cite{Lofwander:2001,Wennerdal:2020}. However, Fig.~\ref{fig:radius_chiral_edge_modes}(c) shows that the spatial dependence is largely independent of subgap energy, which stands in contrast to the ABS. The overall spatial profile of the edge mode LDOS is smooth except for some very small oscillation, but we find that these oscillations reduce monotonically with higher temperature and larger system size, see additional plots in Appendix~\ref{app:additional_results:LDOS}.

Figures~\ref{fig:radius_chiral_edge_modes}(d-l) show that as the disc radius shrinks below $\mathcal{R} < \regimeLengthXi$, significant finite-size effects develop, causing a qualitatively different and highly non-trivial spatial dependence. At $\mathcal{R}=15\xi_0$, small kinks start to develop in the LDOS, see Fig.~\ref{fig:radius_chiral_edge_modes}(d-f). These kinks develop into full peak structures as $\mathcal{R} < 10\xi_0$, see Figs.~\ref{fig:radius_chiral_edge_modes}(g-l). We also find that the typical decay length of the subgap states decreases with $\mathcal{R}$, reaching $\sim2\text{--}3\xi_0$ at $\mathcal{R}=5\xi_0$ (depending on energy), see Figs.~\ref{fig:radius_chiral_edge_modes}(j,l). Hence, the subgap states are ``compressed'' to a smaller spatial region, and in the process obtain a higher magnitude.
Furthermore, as $\mathcal{R}$ decreases, we find both that the largest subgap LDOS peak moves slightly away from the edge and that the gap close to the edge reduces, see in particular Fig.~\ref{fig:radius_chiral_edge_modes}(j-l).

We interpret the finite-size results in Fig.~\ref{fig:radius_chiral_edge_modes} to be due to hybridization between edge states on opposite edges, and due to resonances between the condensate and the edge arising when the system size becomes comparable with the effective coherence length \cite{Wu:2018}. These hybridization effects are naturally enhanced at smaller system sizes, but also with higher temperatures due to a larger effective coherence length.

\section{Spontaneous charge currents}
\label{sec:chiral_ground_state:currents}
In the previous section \ref{sec:chiral_ground_state:edge_modes}, we showed how mesoscopic finite-size effects cause a strong local enhancement of the edge mode LDOS (see Fig.~\ref{fig:radius_chiral_edge_modes}). Here, we study the consequent effects on the chiral charge currents. Specifically, the occupation of the dispersive edge modes leads to a finite charge-current density. It might be tempting to conjecture that since the two chiral branches have roughly the same group velocity they will have a constructive contribution to the total charge current. However, such a simple analysis based on the group velocity can have little bearing on the real situation, and the branches can in fact have the opposite contributions \cite{Wang:2018,Nie:2020}. Consequently, the charge-current density changes direction close to the interface such that the total integrated charge current (and OAM) can vanish. This occurs in particular for all higher-momentum chiral superfluids beyond $p$-wave in the BCS limit and at low temperatures \cite{Huang:2014,Wang:2018}, in a microscopic square lattice \cite{Wang:2018}, but also in some chiral $p$-wave systems \cite{Volovik:2015:b}. The cancellation can be understood in terms of the multiple confinement scales and the current response of surface subgap states versus the condensate \cite{Sauls:2011}, and the contribution from paired versus unpaired fermions \cite{Tada:2015}, see also Ref.~\cite{Nie:2020}.
In contrast, we here show that the mesoscopic finite-size effects studied in the previous sections completely modify the charge-current density, causing instead a strong enhancement of both the total charge current (per edge) and the OMM.

\begin{figure}[t!]
	\includegraphics[width=\columnwidth]{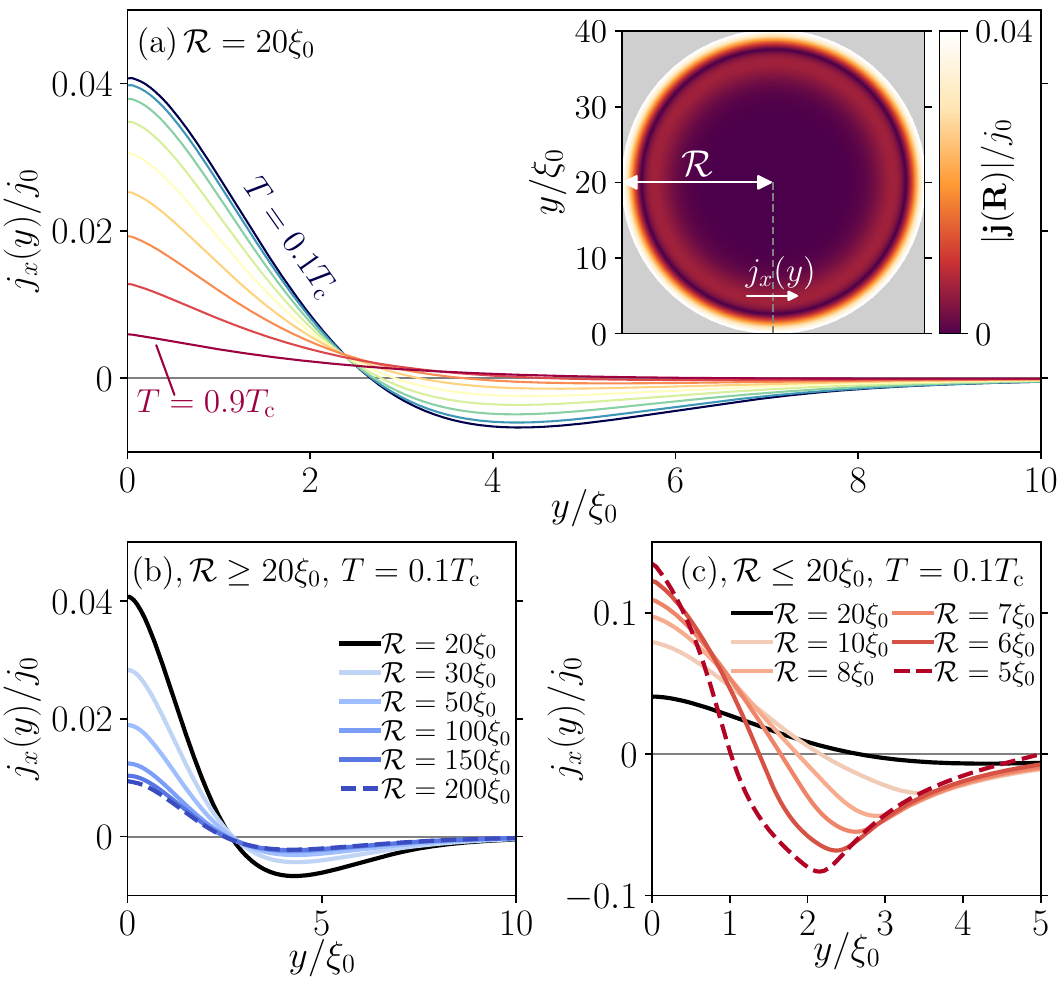}
	\caption{(a) Spatial dependence of the surface-parallel component of the charge-current density $j_x(y)$ in a disc with radius $\mathcal{R}=20\xi_0$ at $\lambda_0=\infty$ with temperatures from $T=0.1\Tc$ (blue) to $T=0.9\Tc$ (red). Horizontal gray line is a guide to the eye marking $j_x=0$. 
        Inset: heatmap of the current-density magnitude at $T=0.1\Tc$. Charge-current density $j_x(y)$ at $T=0.1\Tc$ in systems with radii $\mathcal{R}\geq20\xi_0$ (b) and  $\mathcal{R}\leq20\xi_0$ (c). Note the different ranges of the axes.
        }
	\label{fig:chiral_surface_current}
\end{figure}

\subsection{Charge-current density}
\label{sec:current_density}
We start by analyzing the charge-current density in a chiral $d$-wave superconductor. In Fig.~\ref{fig:chiral_surface_current}(a) we plot the azimuthal (surface-parallel) component of the charge-current density $j_x(y)$ versus perpendicular distance $y$ from the surface, at different temperatures in a system with radius $\mathcal{R}=20\xi_0$. We find an overall monotonic decrease of the magnitude with higher temperatures, which we attribute to thermal suppression of superconductivity. We find two different signs of the charge current as a function of distance from the edge (barely visible at $T=0.9\Tc$), with an overall spatial decay into the bulk over the same characteristic length scale $\decayLength \approx 6\text{--}10\xi_0$ as the order parameter and the edge modes in previous sections.
In Figs.~\ref{fig:chiral_surface_current}(b,c) we analyze the dependence on system size $\mathcal{R}$ by showing the spatial dependence of the charge-current density for $\mathcal{R} \geq \regimeLengthXi$ and $\mathcal{R} \leq \regimeLengthXi$, respectively. In both regimes, there is always a sign change. The current-density magnitude also increases monotonically with decreasing disc radius in agreement with the larger LDOS of the dispersive low-energy edge modes (see Fig.~\ref{fig:radius_chiral_edge_modes}). Apart from these similarities between large and small systems, the exact spatial dependence varies between the two regimes.

In larger systems $\mathcal{R} \geq \regimeLengthXi$ shown in Fig.~\ref{fig:chiral_surface_current}(b), the charge-current density varies smoothly with the distance from the edge, with the sign change occuring at roughly the same distance $y \approx 2.7\xi_0$ for all $\mathcal{R}$.
We find that the overall decay length from the edge remains roughly the same as that of the chiral edge modes ($\decayLength$). As the disc radius increases, the charge-current density reduces monotonically, but seems to saturate around $\mathcal{R} \approx 150\xi_0$. This is similar to the saturation of the small oscillations in the LDOS which also occurs at $\mathcal{R} \approx 150\xi_0$, see the discussion in Sec.~\ref{sec:chiral_ground_state:edge_modes:mesoscopics}. We note that at $\mathcal{R}=20\xi_0$ ($\mathcal{R}=150\xi_0$), the portion of positive and negative charge-current density is very dissimilar (similar) implying a large (small) integrated charge current, discussed further in Sec.~\ref{sec:total_current}. 

In smaller systems $\mathcal{R} \leq \regimeLengthXi$ shown in Fig.~\ref{fig:chiral_surface_current}(c), a reduction of the radius leads to a notable increase in the magnitude of the charge-current density, as well as the sign change occurring at smaller distances, approaching $y\approx 1\xi_0$ at $\mathcal{R}=5\xi_0$.
We point out two important factors contributing to this behavior. The first is the spatial decay of the edge modes themselves. Specifically, as we established in Sec.~\ref{sec:chiral_ground_state:edge_modes:mesoscopics}, a smaller disc radius effectively leads to a ``compression'' of the dispersive edge modes to a smaller region resulting in a higher LDOS magnitude, due to edge-edge hybridization. The occupation of this locally increased LDOS of dispersive states naturally leads to a locally increased charge-current density. The second important factor is that current conservation enforces the charge-current density to sum to zero across the whole system diameter, such that $j_x(y)$ is antisymmetric across the disc center $(x,y)=(\mathcal{R},\mathcal{R})$ where it necessarily goes to zero. For sufficiently small discs this forces $j_x(y)$ to decay to zero below the characteristic decay length $\decayLength$.
As a consequence of these finite-size effects, the portion of positive current becoming smaller but with a larger maximum magnitude, while the portion of negative current increases significantly.

To further supplement the above data, we provide in Appendix~\ref{app:additional_results:currents} the analogue plots of Fig.~\ref{fig:chiral_surface_current}(a) for different $\mathcal{R}$, corroborating that the above analysis holds qualitatively at different temperatures.

In summary, the charge-current density $\vj(\vR)$ in large systems $\mathcal{R}>20\xi_0$ shows the same decay length $\decayLength$ as the edge modes. As $\mathcal{R}$ increases, the spatial profile of $\vj(\vR)$ slowly approaches an asymptotic form. In small systems $\mathcal{R}<20\xi_0$, $\vj(\vR)$ is strongly modified with a shorter decay length but with a much larger magnitude, which we relate to a similar behavior of shorter decay length and larger LDOS magnitude of the dispersive edge modes.

\subsection{Total charge current and magnetic moment}
\label{sec:total_current}
\begin{figure*}[t!]
    \centering
    \begin{minipage}[t]{0.49\textwidth}
	\includegraphics[width=\columnwidth]{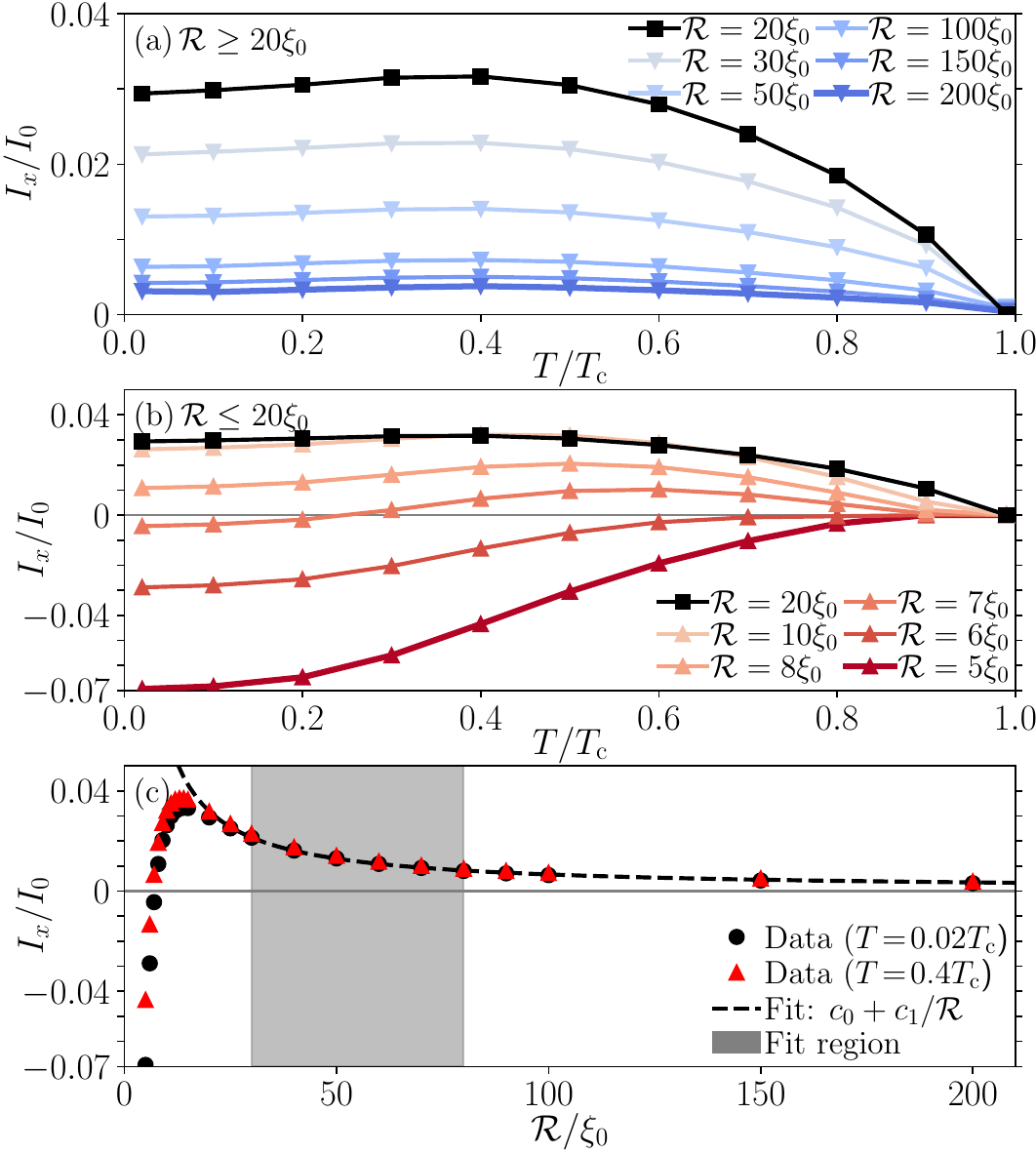}
	\caption{Total charge current $I_x$ [Eq.~(\ref{eq:sheet_current})] versus temperature in large (b) and small (b) systems, and versus system size (c). Markers are calculated data points, solid lines are a guide to the eye, while dashed line is a fit. Here, $\lambda_0=\infty$ and $I_0 \equiv j_0\xi_0 = \hbar|e|\NF\vF^2$. 
        }
	\label{fig:total_current_radius_temperature}
    \end{minipage}\hfill
    \begin{minipage}[t]{0.49\textwidth}
    \includegraphics[width=\columnwidth]{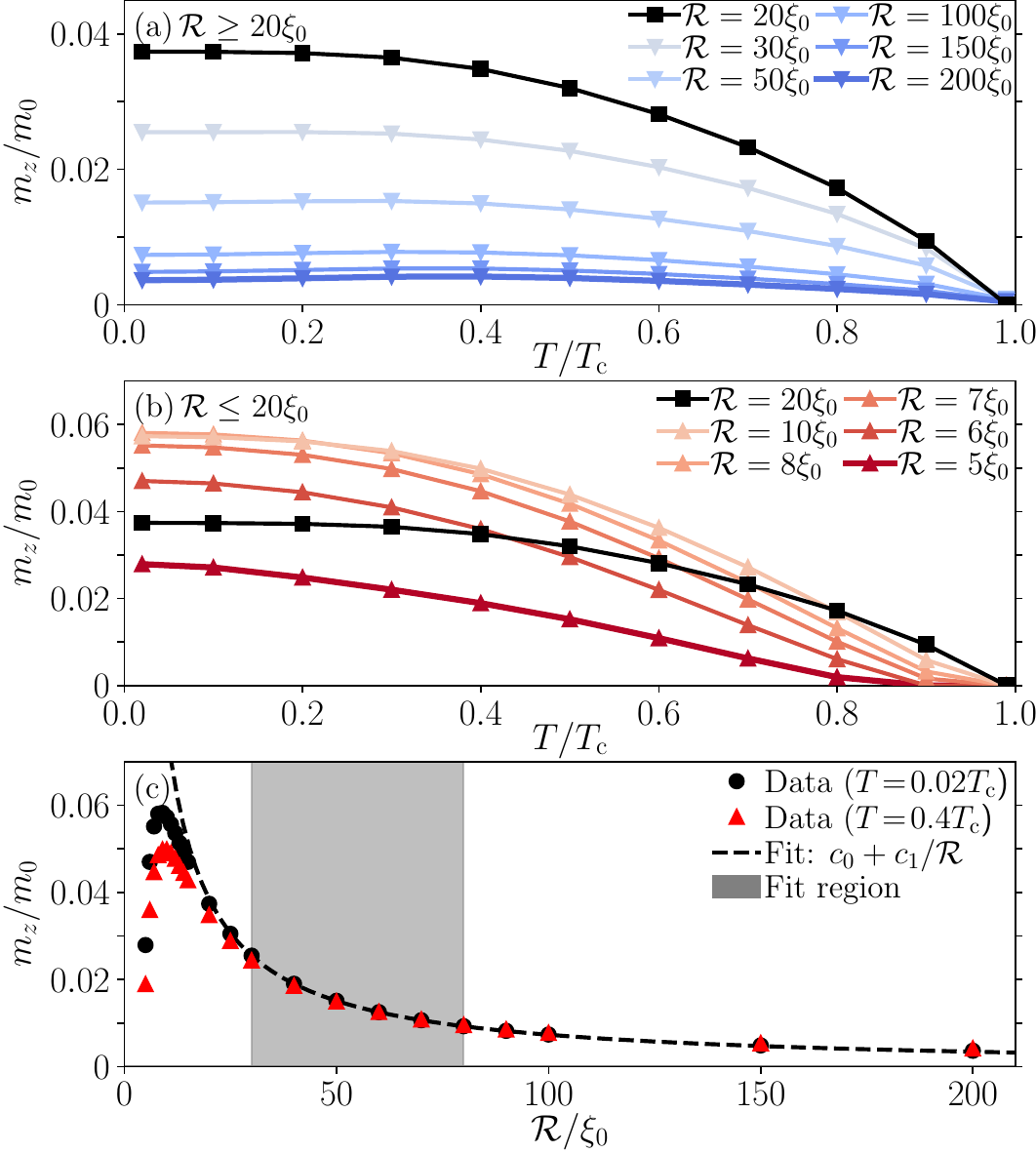}
	\caption{Same as Fig.~\ref{fig:total_current_radius_temperature} but for the OMM $m_z$ [Eq.~(\ref{eq:model:magnetic_moment})]  in units of $m_0 \equiv 2\muB N$.
        }
        \label{fig:magnetic_moment_radius_temperature}
    \end{minipage}
\end{figure*}
We showed in Fig.~\ref{fig:chiral_surface_current} that the charge-current density $\vj(\vR)$ is finite on the length scale of the coherence length. Experimentally resolving signatures of charge currents on such a small scale is already extremely challenging. More feasible techniques are based on measuring e.g.~magnetic signatures of a total charge current averaged over some area. However, the sign-change in the charge-current density reduces such averages. Consequently, the exact balancing of the positive versus negative portions of the charge-current density and its modification by the mesoscopic finite-size effects become important for experimental signatures of the total charge current, and thus for the experimental verification of chiral superconductivity. In this sub-section, we therefore consider the net charge current contribution per edge \footnote{In the finite system studied here the total charge current across the disc diameter is zero (current conservation) $I_x^\prime = \int_{0}^{2\mathcal{R}}j_x(y) dy = 0$ for all $\mathcal{R}$.}, i.e. the total charge current (sheet charge current)
\begin{align}
    \label{eq:sheet_current}
    I_x = \int_{0}^{\mathcal{R}}dy j_x(y),
\end{align}
as a function of disc radius $\mathcal{R}$ and compare against the semi-infinite scenario ($\mathcal{R}\to\infty$) \cite{Wang:2018}. We also consider the OMM $m_z$ [Eq.~(\ref{eq:model:magnetic_moment})] which is averaged over the entire superconducting sample.
Figures~\ref{fig:total_current_radius_temperature} and \ref{fig:magnetic_moment_radius_temperature} show the total charge current $I_x$ and OMM $m_z$, respectively, at $\lambda_0=\infty$. In both figures, (a) and (b) show the temperature dependence at fixed system size $\mathcal{R}$ for the two regimes $\mathcal{R} \geq 20\xi_0$ and $\mathcal{R} \leq 20\xi_0$, respectively, while (c) shows the $\mathcal{R}$ dependence at fixed temperature $T$.

In large systems $\mathcal{R} > 20\xi_0$ shown in Figs.~\ref{fig:total_current_radius_temperature}(a) and \ref{fig:magnetic_moment_radius_temperature}(a), increasing $\mathcal{R}$ monotonically reduces both $I_x$ and $m_z$ at all temperatures $T$, and the reduction shows a slow asymptotic behavior above $\mathcal{R} \approx 150\xi_0$. The reduction is explained by a similar reduction in the magnitude of the charge-current density $\vj(\vR)$ with larger $\mathcal{R}$, and more importantly, that the positive and negative portions of $\vj(\vR)$ become similar in area thus reducing the spatial integral, see Fig.~\ref{fig:chiral_surface_current}(b). The slow asymptotic behavior in $I_x$ and $m_z$ is in turn understood from the similar slow asymptotic behavior in the exact spatial form of $\vj(\vR)$ above $\mathcal{R} \approx 150\xi_0$. Next, we note that $I_x$ has a maximum at $T\approx0.4\Tc$ and a minimum at $T \to 0$ which are both in agreement with calculations in semi-infinite systems \cite{Wang:2018}. Similarly, $m_z$ also seems to develop a slight minimum at low temperature, but only for the largest $\mathcal{R}$. The minima in $I_x$ and $m_z$ tends slowly to zero as $\mathcal{R}\to\infty$, see analysis further below. Most importantly though, both quantities show a dramatic enhancement close to $\mathcal{R} = 20\xi_0$ which is in complete contrast to the semi-infinite scenario, which we explain by the strongly enhanced positive portion of $\vj(\vR)$ [Fig.~\ref{fig:chiral_surface_current}(b)], in turn due to the strong local enhancement of the LDOS of the dispersive edge modes.

In small systems $\mathcal{R} < 20\xi_0$ shown in Figs.~\ref{fig:total_current_radius_temperature}(b) and \ref{fig:magnetic_moment_radius_temperature}(b), $I_x$ has a maximum at $\mathcal{R} \approx 15\text{--}20\xi_0$ while $m_z$ has a maximum at $\mathcal{R} \approx 10\xi_0$. Below these maxima, both quantities decrease monotonically with smaller $\mathcal{R}$. However, this decrease is not described by a decrease in the charge-current density $\vj(\vR)$ since it in fact grows monotonically with smaller $\mathcal{R}$, but is instead completely described by the area becoming similar for its positive and negative portions, see Fig.~\ref{fig:chiral_surface_current}(c).
Interestingly, the negative portion of $\vj(\vR)$ eventually obtains a larger area than the positive portion and starts dominating at low temperatures below $\mathcal{R} < 7\xi_0$, causing a complete reversal of the total charge current, thus suddenly flowing in the opposite direction to the chirality. In contrast, $m_z$ shows no such reversal. These different behaviors in $I_x$ and $m_z$ come from their different integrands in Eqs.~(\ref{eq:sheet_current}) and (\ref{eq:model:magnetic_moment}), specifically $m_z$ always weighs the positive portion of the charge-current density at the edge stronger due to the integrand $\mathbf{R}\times\vj(\vR)$ (note that $\vR$ is the c.m. coordinate with origin in the disc center).

Figures~\ref{fig:total_current_radius_temperature}(c) and \ref{fig:magnetic_moment_radius_temperature}(c) show $I_x$ and $m_z$, respectively, as functions of $\mathcal{R}$ to directly illustrate the scaling behavior. We find that a fit to $f_{\mathrm{fit}}(\mathcal{R}) = c_0 + c_1/\mathcal{R}$ describes the asymptotic behavior for large $\mathcal{R}$ rather well, where $c_0$ and $c_1$ are fit parameters. We choose such a simple model to reduce over-fitting and we verify that fitting only a few data points $30\xi_0 \leq \mathcal{R} \leq 80\xi_0$ (gray region) gives a good extrapolation to all other data points in the interval $\mathcal{R} \geq 20\xi_0$. See Appendix~\ref{app:fit} for further details about the fitting procedure. Importantly, we find that as $T\to0$ and $\mathcal{R}\to\infty$, both $I_x$ and $m_z$ tend to very small values, albeit very slowly. We naturally do not find an exact zero, since our smallest temperature is finite $T = 0.02 \Tc$, we have a finite maximum size $\mathcal{R}=200\xi_0$, our fit model is simple, and due to finite numerical accuracy. Furthermore, we point out that our disc-shaped system has a curved surface in contrast to the straight semi-infinite interface. Thus, even at $\mathcal{R} \approx 150\xi_0$, the surface curves significantly over the distance of a few coherence lengths.
Next, in the opposite limit as $\mathcal{R}$ reduces towards the small regime $\mathcal{R} \leq 20\xi_0$, the $1/\mathcal{R}$ dependence leads to a significantly increased $I_x$ and $m_z$ due to an increase in the positive portion of $\vj(\vR)$, but then a reduction in $I_x$ and $m_z$ as the negative portion of $\vj(\vR)$ starts increasing. As explained previously, this can be directly traced back to the edge-edge hybridization of the dispersive edge modes. Eventually these mesoscopic finite-size effects also cause a breakdown of the $1/\mathcal{R}$ scaling in $I_x$ and $m_z$.

In summary, we find a $1/\mathcal{R}$ scaling with increased system size in both the total charge current $I_x$ and OMM $m_z$. The scaling enhances both quantities in smaller systems, but eventually the scaling breaks down due to the mesoscopic finite-size effects, specifically the edge-edge hybridization and effects of current conservation, causing a local maximum around $\mathcal{R} \approx 20\xi_0$. As a consequence and in contrast to the semi-infinte systems where the current is minimal at low temperatures, we find that both the total charge current and OMM can be large at low temperatures in small systems, pointing to mesoscopic systems as a potential platform to experimentally measure signatures of the chiral currents.

\section{Spontaneous magnetic induction and Meissner screening}
\label{sec:chiral_ground_state:induction}
Any finite charge-current density $\vj(\vR)$, i.e.~like the one we showed in Figs.~\ref{fig:chiral_surface_current}, might induce a magnetic-flux density (induction) according to Eq.~(\ref{eq:model:ampere}) which couples back to the superconductor and generates additional screening currents (self-screening). This spontaneously induced flux serves as a potential experimental observable, measurable with e.g.~nano-SQUIDS and other magnetometry setups \cite{Geim:1997,Bolle:1999,Tsuei:2000,Morelle:2004,Kirtley:2005,Khotkevych:2008,Bleszynski-Jayich:2009,Kokubo:2010,Bert:2011,Jang:2011,Vasyukov:2013,Curran:2014,Kirtley:2016,Ge:2017,BishopVanHorn:2019,Persky:2022}.
Generally, the induction and screening scale inversely with the penetration depth $\lambda_0$ \cite{SuperConga:2023,Holmvall:2017}, tending to zero in the limit $\kappa \equiv \lambda_0/\xi_0 \to \infty$ (the limit studied so far), but may become important when $\lambda_0 < \mathcal{R}$.
It is reasonable to expect that most proposed chiral $d$-wave superconductors fall in the former limit, since most non-elemental or unconventional superconductors are strongly type-II, implying a relatively large penetration depth $\lambda_0$, as is also the case for most thin films and layered materials \cite{Kogan:2014:a,Kogan:2014:b,Ooi:2021,Prozorov:2022}.
However, for sake of completeness, we in this section study the influence of finite penetration depth in the whole type-II interval $\lambda_0\in[1,\infty)\xi_0$, since the penetration depth also depends on materials properties, e.g.~typically increasing (decreasing) with the concentration of non-magnetic (magnetic) impurities in the system \cite{Prozorov:2022}.

\begin{figure}[bt!]
	\includegraphics[width=\columnwidth]{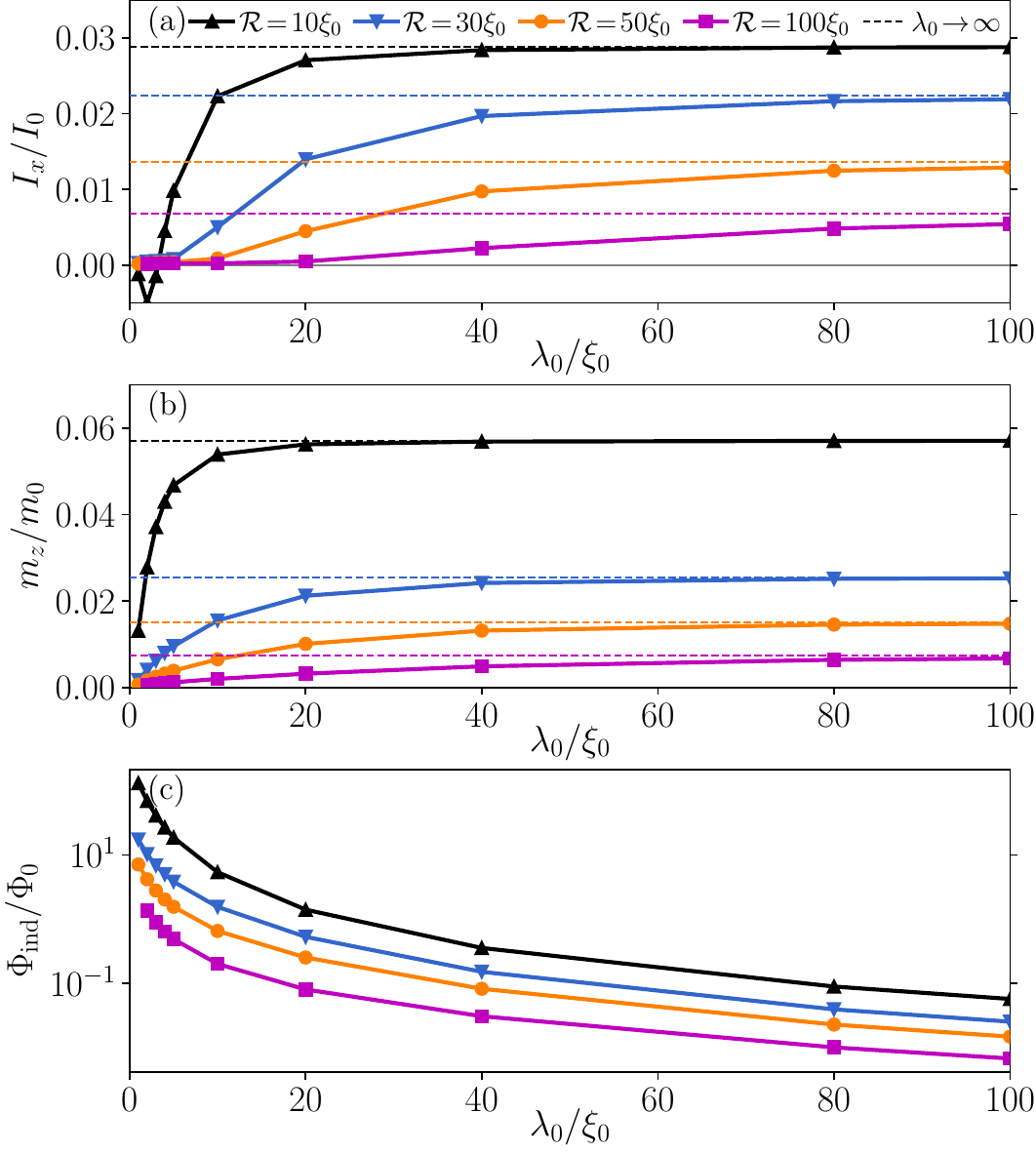}
	\caption{Total charge current $I_x$ (a), OMM $m_z$ (b), total induced magnetic flux $\Phiind$ [Eq.~(\ref{eq:spontaneous_magnetization})] (c), versus finite penetration depth $\lambda_0$ at $T=0.1\Tc$. Note the semi-log scale in (c) in units of flux quantum $\Phi_0\equiv hc/2|e|$. Line colors denote system radius $\mathcal{R}$, while dashed lines denote asymptotes of infinite penetration depth (zero screening).}
	\label{fig:total_current_kappa}
\end{figure}

We begin by summarizing that we find only a small variation in the order parameter profile and in the LDOS for different values of $\lambda_0$ (see Appendices~\ref{app:additional_results:OP} and \ref{app:additional_results:LDOS}, respectively for analysis). We find that the charge-current density $\vj(\vR)$ is unmodified for $\lambda_0 > \mathcal{R}$ but changes significantly as $\lambda_0 \ll \mathcal{R}$ consequently obtaining an additional sign change (see Appendix~\ref{app:additional_results:currents}). The total charge current $I_x$ and OMM $m_z$ show a stronger dependence with $\lambda_0$ as they depend more crucially on the exact balancing between positive and negative portions of $\vj(\vR)$. This we illustrate in Figs.~\ref{fig:total_current_kappa} (a) and (b), respectively, while (c) shows the total induced flux $\Phiind$ defined in Eq.~(\ref{eq:spontaneous_magnetization}).
Quite generally, the ratio between the penetration depth $\lambda_0$ and system size $\mathcal{R}$ essentially determines the importance of the screening \cite{Tinkham:2004}. For $\lambda_0 \gg \mathcal{R}$, we find that the system is poorly screened, with both the total charge current $I_x$ and OMM $m_z$ reaching the asymptotic value of zero screening ($\lambda_0 \to \infty$). For $\lambda_0 < \mathcal{R}$, the screening becomes stronger, and the charge-current density re-organizes to minimize the total charge current in the system. In particular, the chiral charge currents lead to a finite superfluid momentum, which in turn generates a finite kinetic energy. The Meissner screening acts to reduce this kinetic energy. Thus, as $\lambda_0$ becomes small, both $I_x$ and $m_z$ become vanishingly small. This means that in strongly screened systems, the total charge current and OMM is difficult to measure. On the other hand, the charge-current density might still be finite locally for such small $\lambda_0$ (see Appendix~\ref{app:additional_results:currents} for further analysis), leading to an induced magnetic-flux density $\Bind$ via Amp{\`e}re's law in Eq.~(\ref{eq:model:ampere}). Re-writing this equation on a dimensionless form in our natural units $(\xi_0\vnabla)\times\vBind(\vR)/B_0 = (\lambda_0/\xi_0)^{-2}\vj(\vR)/j_0$ with $B_0 \equiv \Phi_0/(\pi\xi_0^2)$, we see that $\Bind$ effectively scales as $\lambda_0^{-2}$ \cite{Holmvall:2017}. Thus, in contrast to $I_x$ and $m_z$, the induced magnetic flux-density $\Bind$ grows dramatically with smaller $\lambda_0$ due to $\vj$ still being finite. This leads to a large total induced magnetic flux $\Phiind$ as we show in Fig.~\ref{fig:total_current_kappa}(c), implying that a signature of the chiral currents can still be measured with e.g.~nano-SQUIDs \cite{Geim:1997,Bolle:1999,Tsuei:2000,Morelle:2004,Kirtley:2005,Khotkevych:2008,Bleszynski-Jayich:2009,Kokubo:2010,Bert:2011,Jang:2011,Vasyukov:2013,Curran:2014,Kirtley:2016,Ge:2017,BishopVanHorn:2019,Persky:2022}.

To summarize, we find that mesoscopic finite-size effects dramatically increase different experimental observables that can be used for verifying chiral superconductivity, specifically the total charge current $I_x$ and OMM $m_z$ for large $\lambda_0$. This scenario is reasonable to expect, since $\lambda_0$ is expected to be large in most proposed chiral $d$-wave superconductors. Even if $\lambda_0$ is for some reason small, our results show that the induced flux $\Phiind$ instead becomes large.

\section{Concluding remarks}
\label{sec:conclusions}
Chiral superconductors are characterized by dispersive chiral edge modes that are presumed to generate spontaneous currents close to interfaces and surfaces. However, with charge not being a conserved quantity in superconductors, the charge current is not topologically protected.
The charge-current density $\vj(\vR)$ may be finite on the superconducting coherence length scale, $\xi_0$, but previous studies have shown that it in many systems also has a sign change close to the surface in all chiral $d$-wave superconductors \cite{Huang:2014,Huang:2015,Tada:2015,Volovik:2015:b,Wang:2018,Nie:2020}. This sign change often causes the total charge current per edge, $I_x$ (i.e.~the integrated charge-current density), to vanish, thereby inhibiting experimental verification of chiral superconductivity.

In this work, we revisit the issue of edge charge currents in chiral $d$-wave superconductors, focusing on mesoscopic sizes relevant for quantum devices, and also computing the OMM $m_z$. We study the dependence of $I_x$ and $m_z$ on the system radius $\mathcal{R}$, temperature $T$, and spontaneous Meissner screening due to a finite penetration depth $\lambda_0$. We show that there are essentially two regimes with qualitatively different behavior, namely large systems $\mathcal{R} > 20\xi_0$ and small systems $\mathcal{R} < 20\xi_0$, with the border inbetween comparable to $\decayLength \approx 10\xi_0$, the characteristic decay (recovery) length of edge (bulk) quantities, e.g.~the order parameter, edge modes, and charge-current density.

For large systems $\mathcal{R} > 20\xi_0$, we show that the total charge current and OMM scale as $1/\mathcal{R}$, recovering previous results of vanishingly small total current as $\mathcal{R} \to \infty$ \cite{Huang:2014,Wang:2018,Nie:2020}.
In smaller systems we find that the $1/\mathcal{R}$ scaling causes a dramatic increase in $I_x$ and $m_z$, but this scaling eventually breaks down below $\mathcal{R} < 20\xi_0$ due to mesoscopic finite-size effects becoming dominant. These effects correspond to current conservation forcing the current to decay faster than $\decayLength$, as well as edge-edge hybridization causing the dispersive edge modes to compress to a smaller region but with a much larger LDOS locally. As a result, the charge-current density $\vj(\vR)$ shows a similar shorter decay length and also with a much larger magnitude locally as $\mathcal{R}$ reduces. Notably, the portions of positive and negative signs in $\vj(\vR)$ re-balance, such that the positive portion closest to the surface shrinks, while the negative portion further away grows. This causes a local maximum to appear in $I_x$ at $\mathcal{R} \approx 15\text{--}20\xi_0$ and in $m_z$ at $\mathcal{R} \approx 10\xi_0$, since they are both described by a spatial integral of $\vj(\vR)$. Importantly, these local maxima survive in the limit of zero temperature, which is in contrast to semi-infinite systems where instead $I_x$ vanishes in the same setup \cite{Wang:2018}. For even smaller systems $\mathcal{R} \lesssim \decayLength$, the re-balancing of the portions of opposite signs in $\vj(\vR)$ eventually causes a sign change in the total charge current, i.e.~the net current spontaneously changes direction due to the mesoscopic finite-size effects, flowing in the opposite direction to the chirality. A reversal of the edge charge current in a chiral superconductor has previously been shown to occur e.g.~due to FS nesting \cite{Bouhon:2014}, surface disorder \cite{Suzuki:2017,Suzuki:2023} or contributions from non-edge states \cite{Bjornson:2015}, while we here show that it also occurs due to mesoscopic finite-size effects.
Furthermore, we show that screening due to finite penetration depth $\lambda_0$ leads to suppression of the charge current and OMM if $\xi_0 \sim \lambda_0 \ll \mathcal{R}$, but we still find an increase in the spontaneously induced magnetic flux $\Phiind$, measurable with magnetometry \cite{Geim:1997,Bolle:1999,Tsuei:2000,Morelle:2004,Kirtley:2005,Khotkevych:2008,Bleszynski-Jayich:2009,Kokubo:2010,Bert:2011,Jang:2011,Vasyukov:2013,Curran:2014,Kirtley:2016,Ge:2017,BishopVanHorn:2019,Persky:2022}. Thus, our results show that mesoscopic finite-size effects significantly enhance the total charge current $I_x$ and OMM $m_z$ in systems with large $\lambda_0$ (relative to the system size), while the induced magnetic flux $\Phiind$ is still significantly large even in systems with small $\lambda_0$.
As a consequence, our results highlight finite mesoscopic systems as a promising platform for enhancing experimental signatures of chiral superconductivity.


\acknowledgements
We thank N.~Wall-Wennerdal, M.~Fogelstr{\"o}m, T.~L{\"o}thman and R.~Arouca for valuable discussions. We acknowledge N.~Wall-Wennerdal, T.~L{\"o}fwander, M.~Fogelstr{\"o}m, M.~H\r{a}kansson, O.~Shevtsov, and P.~Stadler for their work on SuperConga. 
We acknowledge financial support from the Knut and Alice Wallenberg Foundation through the Wallenberg Academy Fellows program. The computations were enabled by the supercomputing resource Berzelius provided by National Supercomputer Centre (NSC) at Link{\"o}ping University and the Knut and Alice Wallenberg foundation. Additional computations were enabled by resources provided by the National Academic Infrastructure for Supercomputing in Sweden (NAISS) and the Swedish National Infrastructure for Computing (SNIC) at C3SE, HPC2N, and NSC, partially funded by the Swedish Research Council through grant agreements No.~2022-06725 and No.~2018-05973. 

\appendix

\section{Riccati formalism}
\label{app:riccati}
Directly solving the Eilenberger equation [Eq.~(\ref{eq:model:eilenberger})] with respect to the propagator $\hat{g}$ typically leads to additional unphysical and unstable solutions. In this Appendix, we show how this can be avoided by expressing the quasiclassical propagators in terms of the Shelankov projectors \cite{Shelankov:1985} which automatically encode the normalization condition. This leads to the so-called Riccati formalism \cite{Shelankov:1985,Nagato:1993,Schopohl:1995,Schopohl:1998,Eschrig:2000,Eschrig:2009,Grein:2013}, which is both more numerically stable and efficient. This formalism is used both in all the numeric simulations, and in all analytic calculations in the subsequent appendices.

The Riccati formalism introduces the coherence amplitudes $\gamma(\vpF,\vR;z)$ and $\tgamma(\vpF,\vR;z)$, sometimes called Andreev amplitudes, as they correspond to probability amplitudes for electron-hole and hole-electron conversion, respectively. Dropping the arguments $(\vpF,\vR;z)$ for brevity, the quasiclassical propagator in Nambu-spin-space is written in terms of $\gamma$ and $\tgamma$ as
\begin{align}
    \label{eq:app:g:riccati}
    \hat{g} & = -i\pi\hat{\mathcal{N}}
    \begin{pmatrix}
        \left(\sigma_0 + \gamma\otimes\tgamma\right) & 2\gamma\\
        -2\tgamma & -\left(\sigma_0 + \tgamma\otimes\gamma\right)
    \end{pmatrix},
\end{align}
with
\begin{align}
    \label{eq:app:N:riccati}
    \hat{\mathcal{N}} & \equiv \begin{pmatrix}
        \left(\sigma_0 - \gamma\otimes\tgamma\right)^{-1} & 0\\
        0 & \left(\sigma_0-\tgamma\otimes\gamma\right)^{-1}
    \end{pmatrix},
\end{align}
where $\otimes$ denotes a matrix product, $\sigma_i$ are the Pauli spin matrices, and the spin dependence is separated into spin-singlet (s) and spin-triplet (t) parts according to
\begin{align}
    \label{eq:app:gamma_spin}
    \gamma & = \left(\gammas + \vgammat\cdot\vsigma\right)i\sigma_2,\\
    \label{eq:app:gamma_tilde_spin}
    \tgamma & = i\sigma_2\left(\tgammas - \vtgammat\cdot\vsigma\right).
\end{align}
By equating Eqs.~(\ref{eq:app:g:riccati})--(\ref{eq:app:N:riccati}) with Eq.~(\ref{eq:model:green_function}), the corresponding spin-space propagators can be derived. Upon substitution into the Eilenberger equation [Eq.~(\ref{eq:model:eilenberger})], a set of coupled ordinary-differential equations are obtained, which are non-linear and of Riccati type. For a spin-singlet superconductor in equilibrium, the equations decouple to two matrix equations in spin-space
\begin{align}
    \label{eq:app:riccati:gamma_singlet:spin}
    \left[i\hbar\vvF \cdot \vnabla + 2z\right]\gamma & = \gamma\tDelta\gamma - \Delta,\\
    \label{eq:app:riccati:gamma_tilde_singlet:spin}
    \left[i\hbar\vvF \cdot \vnabla - 2z\right]\tgamma & = \tgamma\Delta\tgamma - \tDelta.
\end{align}
Spin-degeneracy leads to the scalar Riccati equations
\begin{align}
    \label{eq:app:riccati:gamma_singlet:scalar}
    \hbar\vvF \cdot \vnabla\gammas & = i\Deltas + 2iz\gammas + i\tDeltas\gammas^2,\\
    \label{eq:app:riccati:gamma_tilde_singlet:scalar}
    \hbar\vvF \cdot \vnabla\tgammas & = i\tDeltas - 2iz\tgammas + i\Deltas\tgammas^2,
\end{align}
where additional signs were introduced by terms $(i\sigma_2)^2 = -\sigma_0$. The corresponding scalar propagators take the form
\begin{align}
    \label{eq:app:g:scalar}
    g_0 & = -i\pi\frac{1-\gammas\tgammas}{1+\gammas\tgammas},\\
    \label{eq:app:f:scalar}
    f_{\mathrm{s}} & = -2i\pi\frac{\gammas}{1+\gammas\tgammas}.
\end{align}
and similar for tilde quantities. From here on, we drop the subscript ``s'' and work exclusively with scalar spin-singlet propagators and coherence functions.

We note that the term $\vvF\cdot\vnabla$ which appears in both the Eilenberger and Riccati equations is a directional derivative coupling spatial coordinates with momentum space, since the Fermi velocity $\vvF(\vpF)$ is parametrized by the FS. This term naturally defines a set of ballistic quasiparticle trajectories, where $\gamma$ and $\tgamma$ propagate parallel and antiparallel to $\vvF$, respectively.
It is common to introduce a coordinate system along these trajectories. However, we keep the $xy$-coordinate system in the analytic calculations since we either consider translationally invariant interfaces along which $\partial_x\gamma = 0$ (or equivalently a disc system with large radius of curvature $\mathcal{R} \gg \xi_0$), or bulk systems where $\partial_x\gamma = \partial_y\gamma = 0$. With the definition $\vvF(\vpF) = \vFx(\vpF)\vxhat + \vFy(\vpF)\vyhat$, we explicitly rewrite the operator $\vvF\cdot\vnabla \to \vFx\partial_x + \vFy\partial_y$.

\section{Analytic solution in a uniform environment}
\label{app:analytics:bulk}
This Appendix introduces analytic solutions to a bulk chiral $d$-wave superconductor in the same model as described in Sec.~\ref{sec:model}. These results are used for comparison in the main text against the self-consistent numeric results, as well as in the subsequent Appendix \ref{app:analytics:surface} against analytic results in a semi-infinite system.

In a uniform environment, $\Delta(\vpF,\vR) = \Delta(\vpF)$ and $\vvF\cdot\vnabla\gamma = 0$. The scalar Riccati equations (\ref{eq:app:riccati:gamma_singlet:scalar})-(\ref{eq:app:riccati:gamma_tilde_singlet:scalar}) then take the polynomial form
\begin{align}
    \label{eq:app:riccati:gamma_singlet:homogeneous}
    \tDeltas(\vpF)\gamma^2(\vpF;z) + 2z\gamma(\vpF;z) + \Delta(\vpF) & = 0,\\
    \label{eq:app:riccati:gamma_tilde_singlet:homogeneous}
    \Deltas(\vpF)\tgamma^2(\vpF;z) - 2iz\tgamma(\vpF;z) + \tDelta(\vpF) & = 0.
\end{align}
From these equations, the homogeneous (h) solutions to the Riccati equations are obtained as
\begin{align}
    \label{eq:app:gamma_bulk}
    \gammah(\vpF;z) & = \frac{-\Delta(\vpF)}{z + i\Omega(\vpF;z)},\\
    \label{eq:app:gamma_tilde_bulk}
    \tgammah(\vpF;z) & = \frac{\tDelta(\vpF)}{z + i\Omega(\vpF;z)},
\end{align}
where
\begin{align}
    \label{eq:app:omega}
    \Omega(\vpF;z) & \equiv \sqrt{\Delta(\vpF)\tDelta(\vpF) - z^2}.
\end{align}
For a chiral $d$-wave order parameter, we note that
\begin{align}
    \label{eq:app:delta:bulk:1}
    \Delta(\vpF) & = \Delta_1(\vpF) + i\Delta_2(\vpF)\\
    \label{eq:app:delta:bulk:2}
    & = |\Delta_1|\eta_1(\vpF) \pm i|\Delta_2|\eta_2(\vpF),\\
    \label{eq:app:delta_delta:bulk:1}
    \tDelta(\vpF) & = |\Delta_1|\eta_1(\vpF) \mp i|\Delta_2|\eta_2(\vpF),\\
    \label{eq:app:delta_delta:bulk:2}
    & = \Delta^*(\vpF).
\end{align}
In contrast to the numeric calculations where both amplitudes are allowed to evolve independently, we here assume equal amplitudes appropriate for a degenerate bulk environment
\begin{align}
    \label{eq:app:delta_0}
    |\Delta| \equiv \sqrt{2}|\Delta_1| = \sqrt{2}|\Delta_2|.
\end{align}
Using the trigonometric identity $[\eta_1^2(\vpF) + \eta_2^2(\vpF)]/2 = 1$, we obtain
\begin{align}
    \Delta(\vpF)\tDelta(\vpF) = |\Delta|^2,
\end{align}
Inserting the above solutions and definitions into Eqs.~(\ref{eq:app:g:scalar})--(\ref{eq:app:f:scalar}) yields the bulk propagators
\begin{align}
    \label{eq:app:g:bulk}
    g_0(\vpF;z) & = \frac{-\pi z}{\Omega(\vpF;z)} = \frac{-\pi z}{\sqrt{|\Delta|^2 - z^2}},\\
    \label{eq:app:f:bulk}
    f(\vpF;z) & = \frac{\pi \Delta(\vpF)}{\Omega(\vpF;z)} = \frac{\pi \Delta(\vpF)}{\sqrt{|\Delta|^2 - z^2}},
\end{align}
noting in particular that $g_0(\vpF;z) = g_0(z)$ is independent of $\vpF$ (i.e.~isotropic).

We next use the bulk propagators from Eq.~(\ref{eq:app:g:bulk}) and (\ref{eq:app:f:bulk}) to derive the DOS, the current, as well as the low-temperature gap and free energy. These quantities are used for comparison and as scales throughout the main text. The LDOS is given by Eq.~(\ref{eq:model:ldos}) in terms of the causal (retarded in time) propagator with $z^R = \varepsilon + i\delta$, where $\delta$ is a small positive broadening. Inserting the homogeneous propagator from Eq.~(\ref{eq:app:g:bulk}) and taking the trivial FS average, we get the DOS
\begin{align}
    \label{eq:app:dos_bulk}
    \frac{N(\varepsilon)}{2\NF} = \frac{|\varepsilon|}{\sqrt{\varepsilon^2 - |\Delta|^2}}\Theta\left(\varepsilon^2-|\Delta|^2\right),
\end{align}
i.e.~same as a fully gapped $s$-wave superconductor.

Next for the charge-current density Eq.~(\ref{eq:model:current_density}), we note that the homogeneous Matsubara propagator with $z = i\varepsilon_n$ is given by 
\begin{align}
    \label{eq:app:g_matsubara:bulk}
    g_0^M(\varepsilon_n) & = -\pi\frac{i\varepsilon_n}{\sqrt{|\Delta|^2 + \varepsilon_n^2}},
\end{align}
which is purely imaginary. The current effectively depends on the real part of this propagator \cite{Holmvall:thesis:2019} and is therefore zero. Hence, there are no currents in a uniform chiral $d$-wave superconductor
\begin{align}
    \label{eq:app:j:bulk}
    \vj_{\mathrm{bulk}} = 0.
\end{align}
The breaking of additional symmetries, e.g.~translational symmetry by an interface, can induce chiral edge modes and a finite chiral charge-current density $\vj \neq 0$.

We next turn our attention to the pair propagator, which can be used to solve the low-temperature gap equation in a uniform environment. At low temperatures, we denote the amplitude $|\Delta| \to \Delta_0$, and get the analytic solution
\begin{align}
    \label{eq:app:gap:prefactor:bulk}
    \Delta_0 = \pi e^{-\gammaE}\kB\Tc \approx 1.76 \kB\Tc,
\end{align}
which is the same as for an $s$-wave superconductor \cite{Tinkham:2004} and where $\gammaE$ is the Euler-Mascheroni constant. The low-temperature order parameter is consequently
\begin{align}
    \label{eq:app:op:bulk}
    \Delta(T=0,\vpF) = \frac{\Delta_0}{\sqrt{2}}\left[\eta_1(\vpF) \pm i \eta_2(\vpF)\right],
\end{align}
with $|\Delta(T=0,\vpF)| = \Delta_0$ constant across the whole FS.

Finally, we consider the bulk free energy, i.e. the BCS condensation energy. At low temperature, it can be derived from the Luttinger-Ward functional \cite{Serene:1983,Thuneberg:1983,Vorontsov:2003,Virtanen:2020} and we find it to be
\begin{align}
    \nonumber
    \frac{\Omega_0^{\mathrm{BCS}}}{\mathcal{V}\NF(\kB\Tc)^2} & = -\frac{1}{2} \frac{\langle\Delta(\vpF)\tDelta(\vpF)\rangle_\FS}{(\kB\Tc)^2}\\
    \label{eq:app:free_energy:bulk:2}
    & = \frac{-\Delta_0^2}{2(\kB\Tc)^2}
    \approx -1.56,
\end{align}
which again is the same as for an $s$-wave superconductor \cite{Tinkham:2004}. Here, $\mathcal{V} = \int d^3\vR$ is the volume of the sample.

In summary, we have seen that in a uniform environment, the properties of the chiral $d$-wave superconductor are very similar to that of a conventional $s$-wave superconductor.

\section{Analytic solutions at vacuum interfaces and domain walls}
\label{app:analytics:surface}
In this Appendix, we present analytic solutions to the quasiclassical propagators close to interfaces, taking different boundary conditions into consideration. These solutions are used to interpret the spatial dependence of the numeric results in the main text, in particular in terms of the effective coherence length. Furthermore, we show that perfect specular reflection leads to chiral edge modes and that such a surface is equivalent to a transparent domain wall. Surface retro-reflection and back scattering, on the other hand, lead to a reduced weight of the chiral edge modes. Additionally, the solutions provide a starting point for further analytic studies of the properties of chiral $d$-wave superconductors, such as the thermodynamics, and the spectrum and its full contribution to the angular momentum and equilibrium currents.

\begin{figure}[t!]
	\includegraphics[width=\columnwidth]{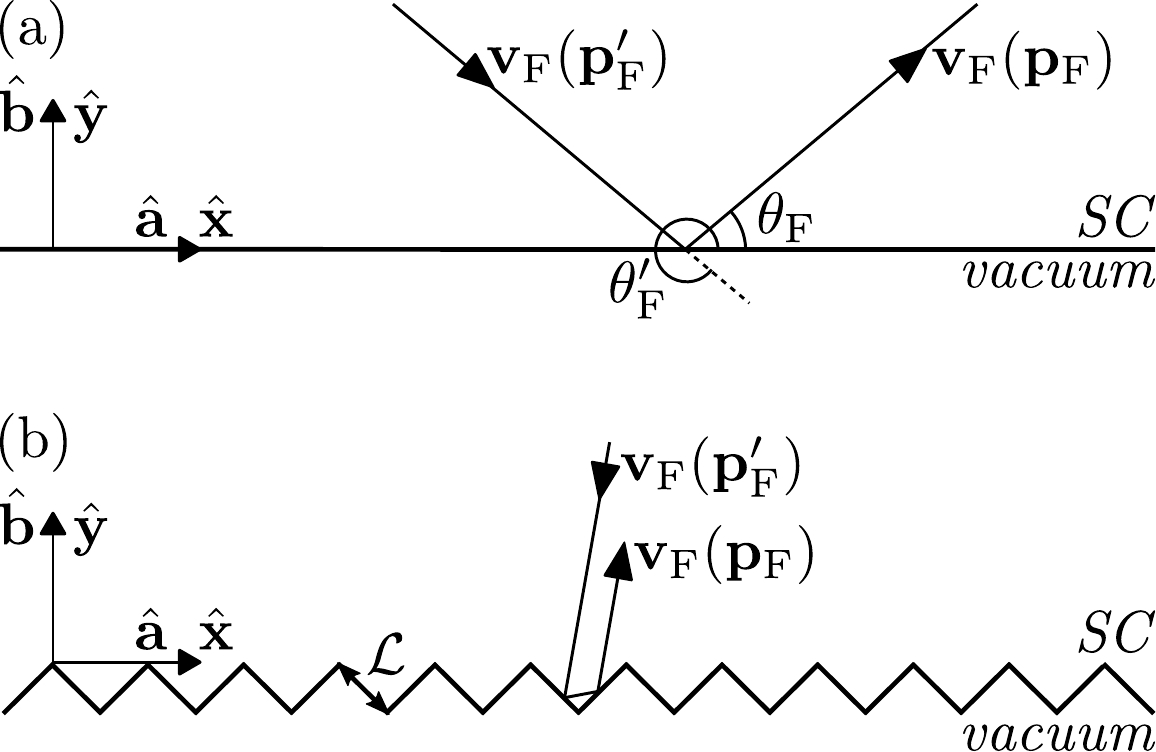}
	\caption{Superconductor-vacuum interface aligned with the crystal $ab$-axes of a semi-infinte superconductor (SC). (a) Translationally invariant interface with perfect specular reflection, connecting incoming (outgoing) quasiparticle momentum $\vpFp$ ($\vpF$) according to $\pFxp = \pFx$ and $\pFyp = -\pFy$. (b) Interface consisting of mesoscopic perfect retro-reflectors of size $\mathcal{L}$ with $a_0 \ll \mathcal{L} \ll \xi_0$, such that surface scattering connects incoming and outgoing quasiparticle momenta according to $\vpFp = -\vpF$.
	}
	\label{fig:app:interface}
\end{figure}

\subsection{Inhomogeneous coherence functions and effective coherence length}
\label{app:analytics:surface:coherence_functions}
We start by deriving the form of the coherence functions in an inhomogeneous environment and show the natural appearance of the effective coherence length discussed in the main text.

Consider a semi-infinite superconductor, as depicted in Fig.~\ref{fig:app:interface}(a). For simplicity, we let the interface normal $y$ be aligned with the high-symmetry axes of the crystal lattice, but note that the orientation does not influence the physics for a chiral order parameter. This is in contrast to e.g.~a nodal $d$-wave superconductor, where $[110]$ and $[100]$ interfaces are distinctly different \cite{Lofwander:2001}. Assuming that any spatial variations in the order parameter can be approximated by a piecewise constant function, the inhomogeneous Riccati equations (\ref{eq:app:riccati:gamma_singlet:scalar})--(\ref{eq:app:riccati:gamma_tilde_singlet:scalar}) can be solved analytically. In the absence of vorticity and other inhomogeneities along the interface coordinate $x$, there is translational invariance $\partial_x\gamma = \partial_x\tgamma = 0$. The only inhomogeneity is then introduced by the boundary, and the inhomogeneous solutions take the explicit form
\begin{align}
    \label{eq:app:riccati:inhomogeneous:gamma}
    \Gamma(\vpF,\vR;z) & = \gammah + \frac{2i\Omega C e^{-y/\xiy}}{1 - \tDelta C e^{-y/\xiy}},\\
    \label{eq:app:riccati:inhomogeneous:gamma_tilde}
    \tGamma(\vpF,\vR;z) & = \tgammah + \frac{2i\Omega \tilde{C} e^{-y/\xiy}}{1 + \Delta \tilde{C} e^{-y/\xiy}},
\end{align}
with integration constants $C$ and $\tilde{C}$, and where the homogeneous solutions $\gammah$ and $\tgammah$, as well as $\Omega$, are given by Eqs.~(\ref{eq:app:gamma_bulk})--(\ref{eq:app:omega}). We note that even in the absence of translational invariance along the interface, the solutions take this form, but with modified exponential factors. Here, $\xiy$ gives the effective superconducting coherence along $y$ (i.e.~perpendicular to the interface)
\begin{align}
    \label{eq:app:xi}
    \xiy(\vpF;z) \equiv \frac{\hbar\left|\vFy(\vpF)\right|}{2\Omega(\vpF;z)},
\end{align}
which can diverge e.g.~at the coherence peaks. We determine the integration constants $C$ and $\tilde{C}$ by imposing continuity wherever there is an inhomogeneitiy, in this case at the boundary of each piecewise constant part in $\Delta$.
This yields the relation $\Gamma(\vpFp,y=0;z) = \Gamma(\vpF,y=0;z)$ and similar for $\tGamma$. Here, $\vpFp = \pF(\cos\thetaFp,\sin\thetaFp)$ with $\thetaFp \in [\pi,2\pi)$ denotes the trajectory impinging on the boundary, while $\vpF = \pF(\cos\thetaF,\sin\thetaF)$ with $\thetaF \in [0,\pi)$ denotes the outgoing trajectory. From here on, we introduce the subscripts ``in'' and ``out'' to denote the impinging and scattered trajectories at the interface, with arguments $(\vpFp,y=0;z)$ and $(\vpF,y=0;z)$, respectively.
This determines the integration constants
\begin{align}
\def\vpF{{\bf p}_{\mathrm{F}}}
    \label{eq:app:C}
    C & = \frac{\Gammain - \gammahout}{2i\Omega_{\mathrm{out}} + \tDelta_{\mathrm{out}}\left(\Gammain - \gammahout\right)},\\
    \label{eq:app:C_tilde}
    \tilde{C} & = \frac{\tGammain - \tgammahout}{2i\Omega_{\mathrm{out}} - \Delta_{\mathrm{out}}\left(\Gammain - \gammahout\right)},
\end{align}
where our notation specifies $\Gammain$ and $\tGammain$ via Eqs.~(\ref{eq:app:riccati:inhomogeneous:gamma}) and (\ref{eq:app:riccati:inhomogeneous:gamma_tilde}), and $\gammahout$ and $\gammahout$ from Eqs.~(\ref{eq:app:gamma_bulk}) and (\ref{eq:app:gamma_tilde_bulk}). From here on, we assume that the important contributions lie in the boundary condition and quasiparticle scattering, rather than in the exact spatial form of the order parameter, and therefore approximate a uniform order parameter $\Delta(\vpF,\vR) \approx \Delta(\vpF)$. This leads to simplification $\Gammain = \gammahin$ and $\tGammain = \tgammahin$. Note that we of course relax this assumptions (and several others) in our numeric simulations. Effects of the spatial dependence of the order parameter on the analytics is discussed e.g. in Ref.~[\onlinecite{Sugiyama:2020}]. We apply the same unitary gauge transformation as in Appendix~\ref{app:analytics:bulk}, such that again $\Delta(\vpF) = \Delta_1(\vpF) + i\Delta_2(\vpF) = |\Delta_1|\eta_1(\vpF) \pm i|\Delta_2|\eta_2(\vpF)$, and $\Delta(\vpF)\tDelta(\vpF) = \Delta\Delta^* = |\Delta|^2$ (i.e. spin-singlet). In the following derivations, we drop the arguments: $(\vpF)$ from $\Delta(\vpF)$, $(\vpF;z)$ from $\Omega(\vpF;z)$, $(\vpF,\vR;z)$ from $\gamma(\vpF,\vR;z)$, and similar for tilde quantities.

\subsection{Specular boundary conditions: reflective vacuum interfaces and transparent domain walls}
\label{app:analytics:surface:specular}
We next proceed to derive the surface propagators and their spatial dependence at interfaces with perfect specular boundary conditions, followed by expressions for the chiral edge modes and chiral edge currents. These expressions form a base line to compare against fully self-consistent results in the main text.

Perfect specular scattering at an interface is characterized by the momentum change $(p_\parallel,p_\perp) \to (p_\parallel,-p_\perp)$, connecting two trajectories with angles $\thetaFp + \thetaF = 2\pi$, as shown in Fig.~\ref{fig:app:interface}(a). In both the chiral $p$-wave and $d$-wave systems, this leads to a sign-change in one of the two order parameter components, while the other is uninfluenced. In our notation, this means that from an incoming to an outgoing trajectory, $(\Delta_1,\Delta_2) \to (\Delta_1,-\Delta_2)$, and consequently $\Delta \to \Delta^*$. In other words, a quasiparticle undergoing such a scattering sees a chiral inversion, in exactly the same way as if it transmits through a transparent domain wall. The two scenarios are treated analogously. This leads to
\begin{align}
    \label{eq:app:specular:gamma:continuity}
    \Gammain - \gammahout & = -\frac{\Delta^* - \Delta}{z+i\Omega} = -\frac{2i\Delta_2}{z+i\Omega},\\
    \label{eq:app:specular:gamma_tilde:continuity}
    \tGammain - \tgammahout & = \frac{\tDelta^* - \tDelta}{z+i\Omega} = -\frac{2i\Delta_2}{z+i\Omega}.
\end{align}
Inserting these solutions into Eqs.~(\ref{eq:app:C})--(\ref{eq:app:C_tilde}) yields the integration constants
\begin{align}
\def\vpF{{\bf p}_{\mathrm{F}}}
    \label{eq:app:C:specular}
    C & = \frac{\Delta_2}{\Omega(z+i\Omega) + \tDelta\Delta_2},\\
    \label{eq:app:C_tilde:specular}
    \tilde{C} & = \frac{\Delta_2}{\Omega(z+i\Omega) - \Delta\Delta_2}.
\end{align}
such that after a bit of algebra, we find the inhomogeneous solutions
\begin{align}
    \label{eq:app:Gamma:specular}
    \Gamma & = \gammah\frac{\Omega\Delta + (z-i\Omega)\Delta_2 - (z+i\Omega)\Delta_2 e^{-y/\xiy}}{\Omega\Delta + (z-i\Omega)\Delta_2 - (z-i\Omega)\Delta_2 e^{-y/\xiy}},\\
    \label{eq:app:Gamma_tilde:specular}
    \tGamma & = \tgammah\frac{\Omega\tDelta - (z-i\Omega)\Delta_2 + (z+i\Omega)\Delta_2 e^{-y/\xiy}}{\Omega\tDelta - (z-i\Omega)\Delta_2 + (z-i\Omega)\Delta_2 e^{-y/\xiy}}.
\end{align}
Using these solutions, we can solve for the surface propagators via Eqs.~(\ref{eq:app:g:scalar}) and (\ref{eq:app:f:scalar}), but note that we have to treat the direction of $\vvF(\vpF)$ carefully. For $\thetaF \in [0,\pi)$ we get $\gammas = \Gamma$ and $\tgammas = \tgammah$, while we for $\thetaF \in [\pi,2\pi)$ we get $\gammas = \gammah$ and $\tgammas = \tGamma$. We find that the solutions can be generalized for all $\thetaF$ by introducing the terms
\begin{align}
    \label{eq:app:s_factor}
    s = \operatorname{sgn}(\vFy).
\end{align}
The surface propagators then become
\begin{align}
    \label{eq:app:g:specular}
    g_0(\vpF,\vR;z) = & -\frac{\pi z}{\Omega} -\frac{\pi s \Delta_2}{\Omega}\frac{\Omega\Delta_1 - z s \Delta_2}{z^2 - \Delta_1^2}e^{-y/\xiy},\\
    \nonumber
    f(\vpF,\vR;z) = & \frac{\pi i\Delta_2}{\Omega}\left(1 - e^{-y/\xiy}\right) + \frac{\pi\Delta_1}{\Omega} \\
    \label{eq:app:f:specular}
    & + \frac{\pi s\Delta_2}{\Omega}\frac{\Delta_2^2 - z^2}{\Omega z + s\Delta_1\Delta_2}e^{-y/\xiy},
\end{align}
which are valid at a specular reflective interface and at a transparent domain wall. We see that far from the interface, where $e^{-y/\xiy} \to 0$, the bulk propagators in Eqs.~(\ref{eq:app:g:bulk})-(\ref{eq:app:f:bulk}) are recovered. Close to the surface ($e^{-y/\xiy} \to 1$), the propagators take the form
\begin{align}
    \label{eq:app:g:specular:surface}
    g_0(\vpF,y=0;z) = & -\frac{\pi z}{\Omega} -\frac{\pi s \Delta_2}{\Omega}\frac{\Omega\Delta_1 - z s \Delta_2}{z^2 - \Delta_1^2},\\
    \label{eq:app:f:specular:surface}
    f(\vpF,y=0;z) = & \frac{\pi\Delta_1}{\Omega} + \frac{\pi s\Delta_2}{\Omega}\frac{\Delta_2^2 - z^2}{\Omega z + s\Delta_1\Delta_2},
\end{align}
We next take a closer look at the edge modes and currents related to these surface propagators.

\subsubsection{Chiral edge modes}
\label{app:analytics:surface:edge_modes}
The propagators in Eqs.~(\ref{eq:app:g:specular})--(\ref{eq:app:f:specular}) have a simple pole at $z = s\Delta_1$, seen via the residue
\begin{align}
    \label{eq:app:residue_plus}
    \operatorname{Res}[g(z=s\Delta_1)] & = \pi \left|\Delta_2\right|e^{-y/\xi_y}.
\end{align}
This corresponds to bound states at energy $\varepsilon_{\mathrm{bs}} = s\Delta_1$ which is a propagating chiral mode with momentum $\pFx = \pF \cos(\thetaF)$. The spectrum of the chiral edge modes can essentially be derived via the momentum-resolved LDOS, i.e. omitting the angular average in Eq.~(\ref{eq:model:ldos}). This yields the spectrum plotted in Fig.~\ref{fig:chiral_spectrum}(a), thus with two edge modes for $s=\pm1$. 

\subsubsection{Chiral surface currents}
\label{app:analytics:surface:currents}
In equilibrium, the Fermi-Dirac distribution together with the chiral edge modes lead to the occupation of a state with a net momentum, breaking time-reversal symmetry and generating a charge-current density. We next study this charge-current density, defined by Eq.~(\ref{eq:model:current_density}), in terms of the Matsubara propagators $g^{\mathrm{M}}(\vpF,\vR;\varepsilon_n) = g_0(\vpF,\vR;z=i\varepsilon_n)$ with Matsubara energies $\varepsilon_n = \pi\kB T(2n+1)$. We find that the real part of the Matsubara quasiparticle propagator is
\begin{align}
    \label{eq:app:g_matsubara}
    \operatorname{Re}\left[g^{\mathrm{M}}(\vpF,\vR;\varepsilon_n)\right] = \pi s \frac{\Delta_1\Delta_2}{\Delta_1^2 + \varepsilon_n^2},
\end{align}
such that the chiral edge current is given by
\begin{align}
    \label{eq:app:chiral_current}
    \frac{j_x(y)}{j_0} = -2\frac{T}{\Tc} \left\langle \vFx s\Delta_1\Delta_2 \sum_{\varepsilon_n > 0}^{\Omegac} \frac{e^{-y/\xiy}}{\Delta_1^2 + \varepsilon_n^2} \right\rangle_\FS.
\end{align}
Here, the overall sign is set by that of $\Delta_2$, i.e. the chirality directly sets the direction of the current. The above current shows a sign change at $\sim 1\xi_0$, similar to the fully self-consistent numerical calculations in Fig.~\ref{fig:chiral_surface_current}. However, the overall magnitude differs significantly, which we interpret to be due to the approximation in the spatial dependence of the order parameter \cite{Sugiyama:2020}, and other non-trivial effects omitted in the analytical but not in the numerical calculations.

\subsection{Retro-reflection and back scattering}
\label{app:analytics:surface:retro_reflection}
We next study interfaces with retro-reflection and back scattering (defined below), and show that it is an interesting scenario that can be used to further distinguish between spin-singlet and spin-triplet superconductors, e.g.~between chiral $d$-wave and chiral $p$-wave states. We note that retro-reflection becomes experimentally relevant even in superfluid ${{^{3}}\text{He}}$ \cite{Heikkinen:2021}.

Surface roughness and diffusivity causes a finite probability of back scattering. In the extreme limit of perfect retro-reflection, e.g.~at a surface with a mesoscopic saw-tooth profile of perfect reflectors with length $\mathcal{L}$ (where $a_0 \ll \mathcal{L} \ll \xi_0$) as depicted in Fig.~\ref{fig:app:interface}(b), then $(p_\parallel,p_\perp) \to (-p_\parallel,-p_\perp)$. A spin-singlet order parameter is symmetric with respect to such a momentum reversal, $\Delta(-\vpF) = \Delta(\vpF)$, which leads to $\Gammain = \gammahout$, and consequently $C=\tilde{C} = 0$, such that
\begin{align}
    \label{eq:app:gamma:retro}
    \Gamma_{\mathrm{retro}}(\vpF,\vR;z) & = \gammah(\vpF;z),\\
    \label{eq:app:gamma_tilde:retro}
    \tGamma_{\mathrm{retro}}(\vpF,\vR;z) & = \tgammah(\vpF;z).
\end{align}
Thus, despite the presence of the interface, this yields the same bulk propagators and solutions presented in Appendix~\ref{app:analytics:bulk}. In contrast, a spin-triplet system is instead antisymmetric $\Delta(-\vpF) \to -\Delta(\vpF)$, and the retro-reflective boundary condition therefore leads a Jackiw-Rebbi zero mode \cite{Jackiw_Rebbi:1976} associated with surface pairbreaking and the appearance of flat bands of ABS. Hence, perfect retro-reflection technically causes a removal of the chiral edge modes since quasiparticles essentially scatter between the same Chern numbers, but this result is highly unstable as any infinitesimal modification from perfect retro-reflection leads to a recovery of the chiral edge modes, see also Ref.~\cite{Sauls:2011} for further discussion. Thus, for more realistic back scattering, there is instead a finite suppression of the LDOS of the chiral edge modes, in favor of a bulk (flat-band ABS) LDOS in spin-singlet (spin-triplet) systems.

The above results point to an interesting scenario to distinguish between spin-singlet and spin-triplet order parameters, since the spin-singlet system effectively behaves like a bulk system close to a retro-reflecting interface, while the spin-triplet system leads to order parameter suppression and a large density of dispersionless surface ABS with energies at the Fermi level.

\section{Additional numerical results}
\label{app:additional_results}
This appendix presents additional numerical results for a chiral $d$-wave superconductor to supplement the analysis in the main text. The first subsection studies the order parameter, the second subsection studies the LDOS of the zero-energy edge modes, while the third subsection studies the charge-current density.

\subsection{Order parameter}
\label{app:additional_results:OP}
\begin{figure*}[t!]
	\includegraphics[width=\textwidth]{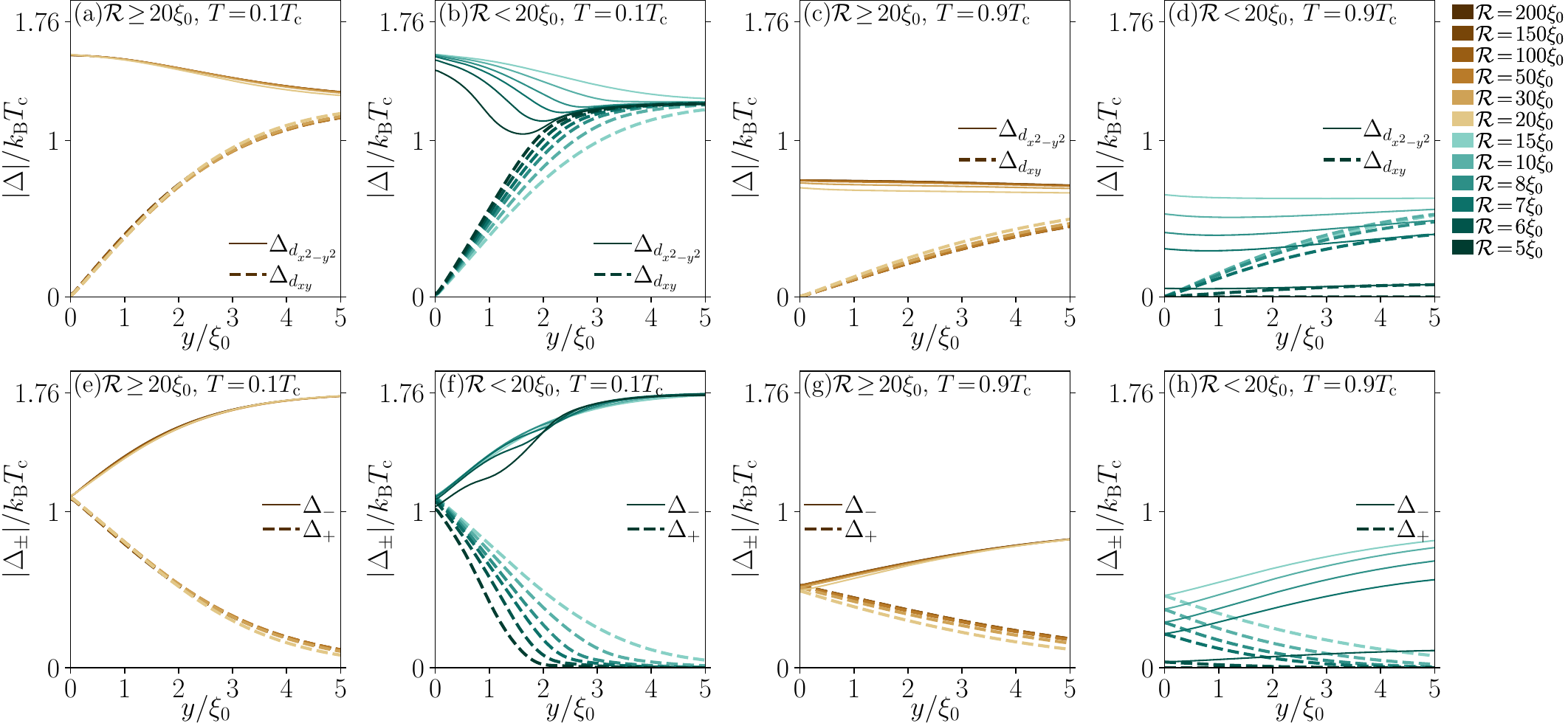}
	\caption{Order parameter amplitudes as a function of distance from a $[100]$ edge in a disc with radius $\mathcal{R}$ and dominant chirality $\Dminus$ at $\lambda_0=\infty$, for the nodal components (top row) and chiral components (bottom row). First (second) column: low temperature in large (small) discs. Third (fourth) column: high temperature in large (small) discs.}
	\label{fig:app:chiral_OP:radius}
\end{figure*}
\begin{figure}[t!]
	\includegraphics[width=\columnwidth]{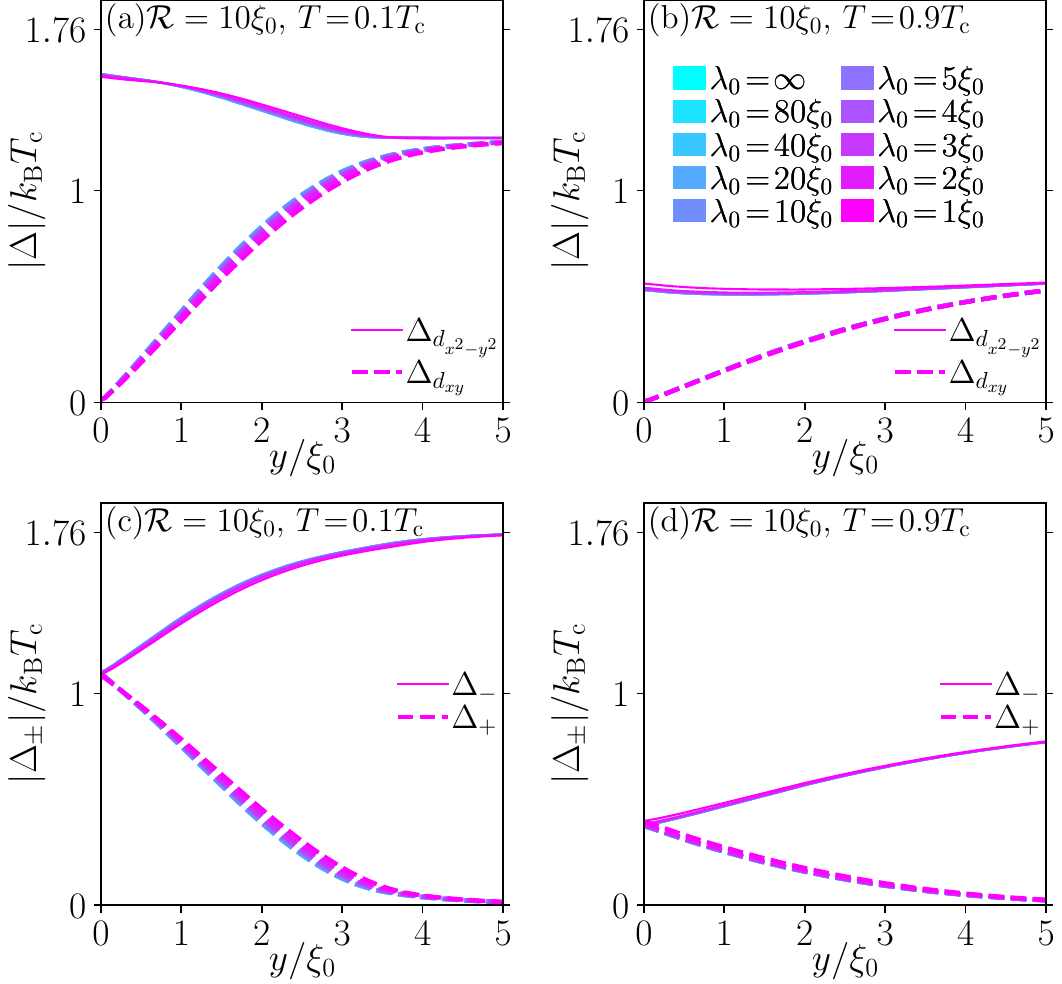}
	\caption{Order parameter amplitudes as a function of distance from a $[100]$ edge in a disc with dominant chirality $\Dminus$, for the nodal components (top row), and chiral components (bottom row). First (second) column: low (high) temperature.}
	\label{fig:app:chiral_OP:kappa}
\end{figure}

We next study how the spatial dependence of the order parameter is influenced by mesoscopic finite-size effects in Fig.~\ref{fig:app:chiral_OP:radius} and by finite Meissner screening in Fig.~\ref{fig:app:chiral_OP:kappa}, to be compared with the analogous Fig.~\ref{fig:chiral_ground_state:op} in the main text.

Figure~\ref{fig:app:chiral_OP:radius} shows the amplitude of the nodal (chiral) order parameter components in the top (bottom) row, as a function of perpendicular distance $y$ from a $[100]$ interface, in a disc-shaped superconductor with radius $\mathcal{R}$, dominant chirality $\Dminus$, and penetration depth $\lambda_0=\infty$. The first (second) column shows large (small) discs at low temperature $T=0.1\Tc$, while the third (fourth) column shows large (small) discs at high temperature $T=0.9\Tc$. The nodal component $\Dxtyt$ ($\Dxy$) is enhanced (suppressed) due to fermionic bound states at the surface \cite{Sauls:2011}, but recover to their bulk values over a characteristic length $\decayLength$, i.e.~the same as in the main text, which is also the decay length of the bound states. Similarly, the dominant (subdominant) chiral component $\Dminus$ ($\Dplus$) is suppressed (enhanced) at the edge, but also recover to their bulk value over the length scale $\decayLength$. In large systems $\mathcal{R} > 20\xi_0$, the order parameter profiles remain the same for all disc radii, with the same $\decayLength \approx 6\text{--}10\xi_0$ (depending on e.g.~temperature). As $\mathcal{R}$ becomes comparable with $\decayLength$, the influence of opposite edges becomes important, leading to hybridization as discussed in Sec.~\ref{sec:chiral_ground_state:edge_modes:mesoscopics}. Therefore, in small systems $\mathcal{R} < 20\xi_0$, the order parameter profiles vary strongly with disc radius, with the typical decay length reducing monotonically to $\sim 2\xi_0$ at $\mathcal{R}=5\xi_0$. Finally, we note that increased temperatures lead to an overall suppression of the order parameter due to thermal excitations, which leads to a longer effective coherence length as discussed in Sec.~\ref{sec:chiral_groundstate}. Hence, finite-size effects are significantly enhanced at higher temperatures. This leads to a complete destruction of superconductivity at $T=0.9\Tc$ in the smallest disc with $\mathcal{R}=5\xi_0$. Thus, the above results show a strong dependence of the exact order parameter profile with system size.

Figure~\ref{fig:app:chiral_OP:kappa} shows the spatial dependence of the order parameter amplitude in a disc with radius $\mathcal{R}=10\xi_0$, for different $\lambda_0$, illustrating negligible influence for all $\lambda_0 \in [1,\infty)\xi_0$ at both high and low temperatures. The influence is even smaller for larger $\mathcal{R}$ (not shown). Hence, the self-screening effects due to small $\lambda_0$ discussed in Sec.~\ref{sec:chiral_ground_state:induction} does not correspond to a modified order parameter amplitude, but rather the subgap states and the exact shape of the charge-current density as discussed next.

\subsection{Local density of states}
\label{app:additional_results:LDOS}
\begin{figure}[t!]
	\includegraphics[width=\columnwidth]{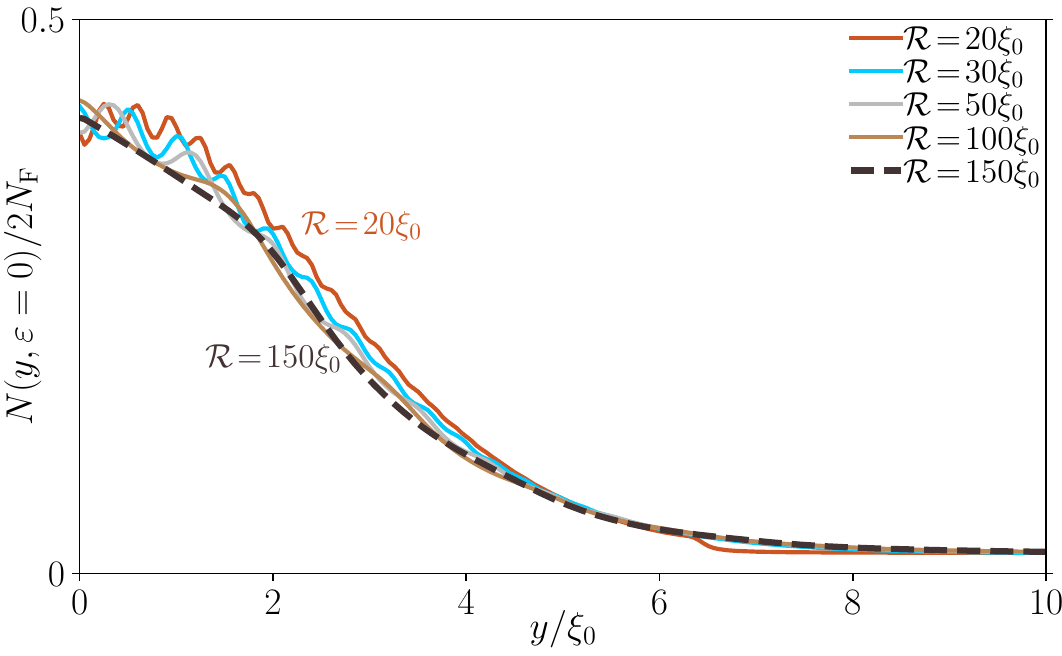}
	\caption{Spatial dependence of the edge modes at zero energy, $T=0.5\Tc$ and $\lambda_0=\infty$, for different disc radii $\mathcal{R}$.}
	\label{fig:app:ldos_vs_R}
\end{figure}

\begin{figure}[t!]
	\includegraphics[width=\columnwidth]{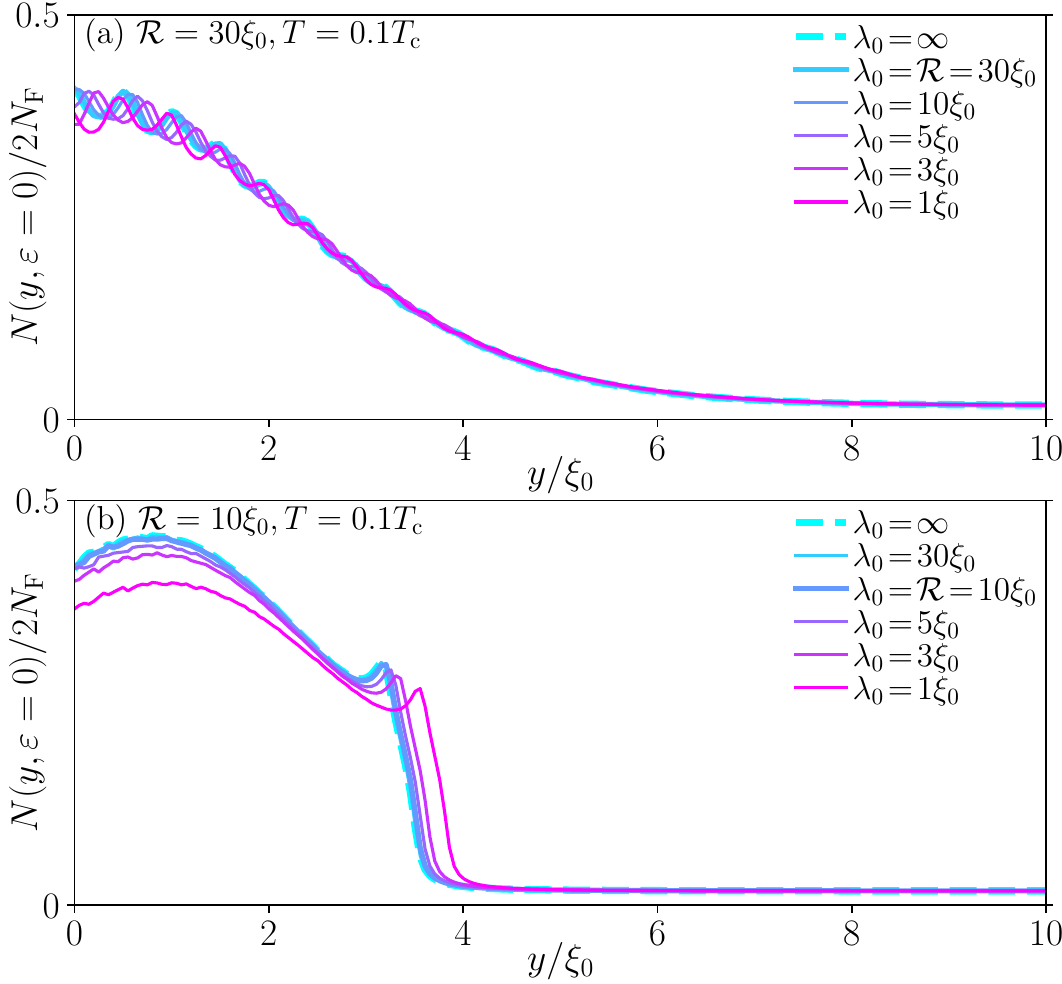}
	\caption{Spatial dependence of the edge modes at zero energy and $T=0.1\Tc$ in the large disc regime $\mathcal{R}=30\xi_0$ (a) and small disc regime $\mathcal{R}=10\xi_0$ (b), from $\lambda_0=1\xi_0$ (pink) to $\lambda_0=\infty$ (cyan).}
	\label{fig:app:ldos_vs_kappa}
\end{figure}

We next study how the spatial dependence of the zero-energy LDOS approaches an asymptotic form for large disc radius $\mathcal{R}$ in Fig.~\ref{fig:app:ldos_vs_R}, and how the spatial dependence is influenced by finite Meissner screening in Fig.~\ref{fig:app:ldos_vs_kappa}, to be compared with e.g.~the bottom row in Fig.~\ref{fig:radius_chiral_edge_modes} in the main text.

Figure~\ref{fig:app:ldos_vs_R} plots the LDOS at zero energy as a function of distance from the edge, for different disc radii $\mathcal{R}$, at fixed $T=0.5\Tc$ and $\lambda_0=\infty$. Importantly, the spatial decay length $\decayLength \approx {6\text{--}10\xi_0}$ is roughly the same in all large systems $\mathcal{R} > 20\xi_0$ and there are no strong kinks or additional peak structures like in the small systems $\mathcal{R} \leq 20\xi_0$. Hence, the overall spatial dependence is quite smooth, apart from a small oscillations. These oscillations reduce significantly with larger $\mathcal{R}$ and subsides almost completely at $\mathcal{R} \approx 150\xi_0$. We note that this is the size where the charge-current density reaches an asymptotic value, see Fig.~\ref{fig:chiral_surface_current}(b). However, it is not clear at the present if this similar asymptotic behavior with large $\mathcal{R}$ in the edge modes and charge-current density are related, or if this is more of a coincidence.

Figure~\ref{fig:app:ldos_vs_kappa} shows the spatial dependence of the zero-energy edge modes at $T=0.1\Tc$ but for different penetration depths, in a large disc $\mathcal{R}=30\xi_0$ (a), and a small disc $\mathcal{R}=10\xi_0$ (b). In the large disc, different $\lambda_0$ mainly influence the small oscillations in the LDOS, but the edge mode decay is otherwise not influenced. In the small disc, a reduction of $\lambda_0$ causes a smaller LDOS magnitude and slight increase in the decay length from the edge. Since the LDOS corresponds to dispersive edge modes, these results suggest that a stronger screening reduces the currents, which is consistent with the results found in the main text.

\subsection{Currents}
\label{app:additional_results:currents}
\begin{figure*}[bt!]
	\includegraphics[width=\textwidth]{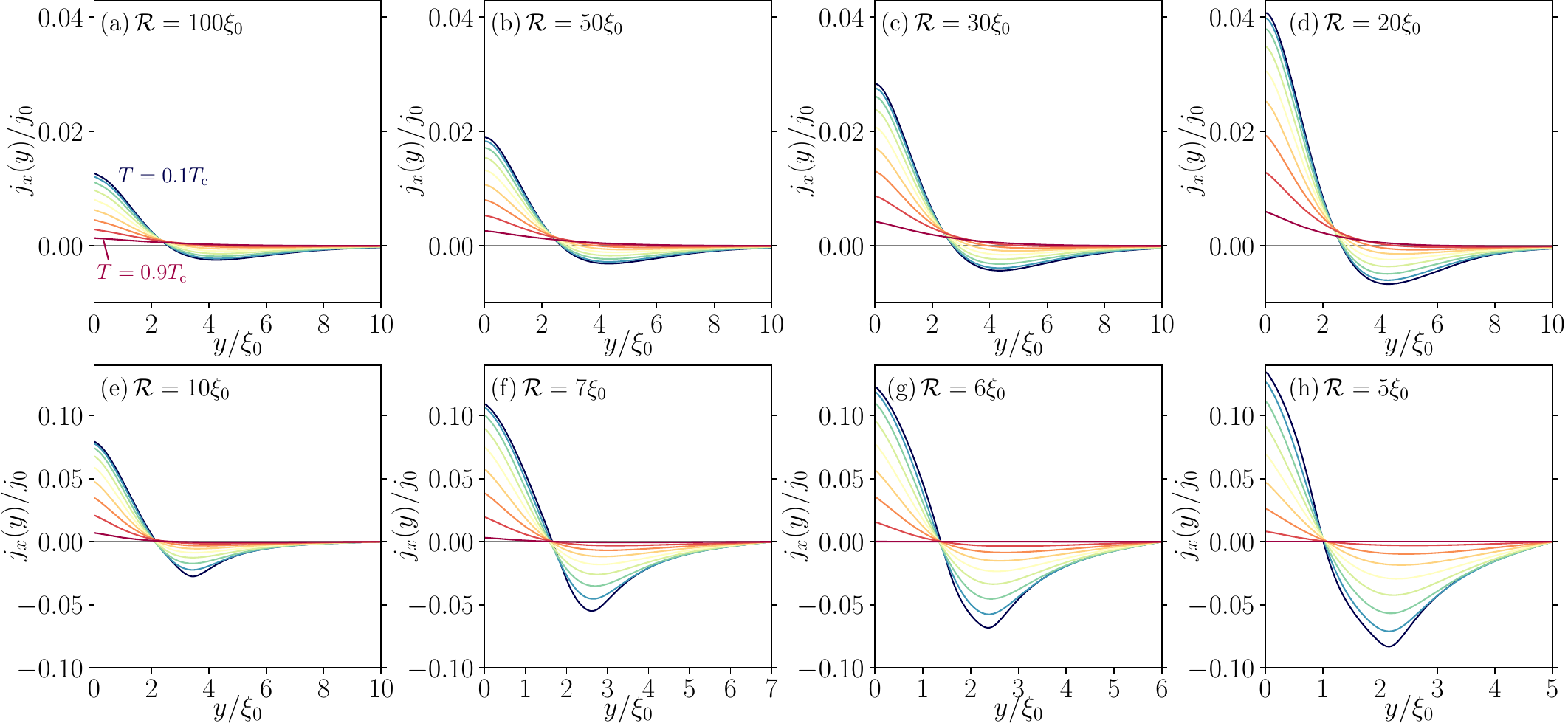}
	\caption{Azimuthal (surface-parallel) component of the charge-current density as a function of distance from the edge in a disc with dominant chirality $\Dminus$ at $\lambda_0=\infty$. Different panels correspond to different $\mathcal{R}$, with temperatures $T=0.1{\Tc}$ (blue) to $T=0.9{\Tc}$ (red). Horizontal line (gray) marks $j_x=0$. Note different $x$- and $y$-axis scales in top versus bottom rows.}
	\label{fig:app:chiral_currents_radius}
\end{figure*}

\begin{figure}[bt!]
	\includegraphics[width=\columnwidth]{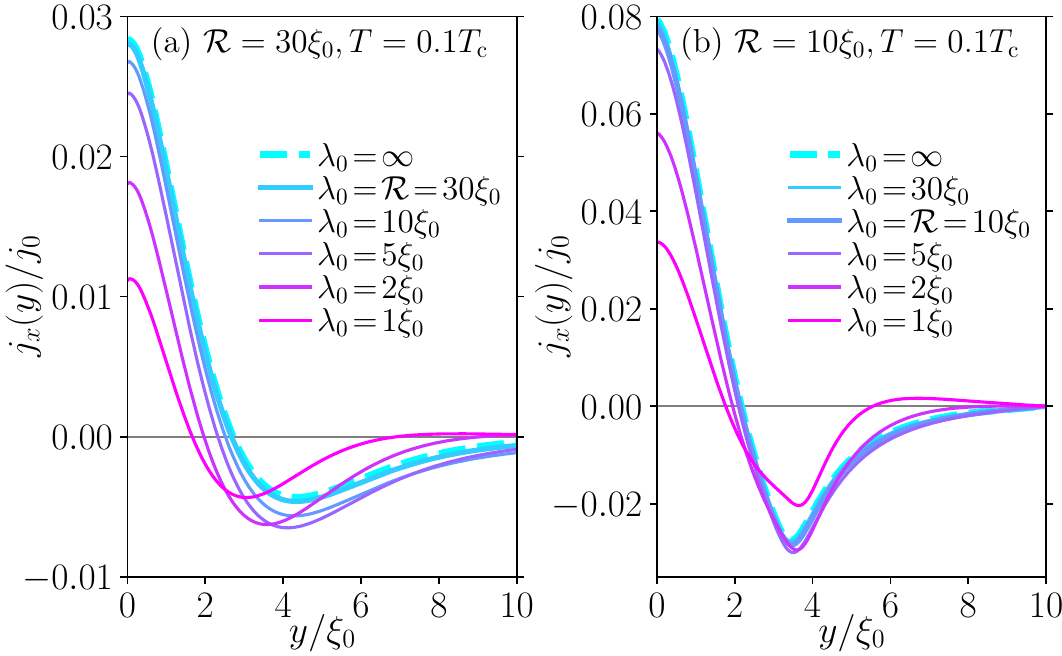}
	\caption{Spatial dependence of the azimuthal (surface-parallel) component of the charge-current density at $T=0.1\Tc$ in a disc with radius $\mathcal{R}=30\xi_0$ (a) and $\mathcal{R}=10\xi_0$ (b), for different $\lambda_0$ from $\lambda_0=1\xi_0$ (pink) to $\lambda_0=\infty$ (cyan). Horizontal line (gray) marks $j_x = 0$.
        }
	\label{fig:chiral_currents_kappa}
\end{figure}

We next study how the spatial dependence of the charge-current density depends on temperature for different disc radii $\mathcal{R}$ in Fig.~\ref{fig:app:chiral_currents_radius}, i.e.~the direct analogue of Fig.~\ref{fig:chiral_surface_current}(a). We also study how the charge-current density is influenced by finite screening in Fig.~\ref{fig:chiral_currents_kappa}, supplementing the analysis in Sec.~\ref{sec:chiral_ground_state:induction}.

Figure~\ref{fig:app:chiral_currents_radius} shows the spatial dependence of the charge-current density $\vj(\vR)$ at $\lambda_0=\infty$ and at different temperatures, where each panel corresponds to a different system radius $\mathcal{R}$. Below $\mathcal{R} \leq 10\xi_0$ the system radius is smaller than the typical decay length $\decayLength$ of the currents. This leads to a significantly increased magnitude due to larger LDOS of dispersive edge modes as discussed in Sec.~\ref{sec:chiral_ground_state:edge_modes:mesoscopics}, and a more complicated spatial dependence with kink-like structures (higher-order harmonics) due to the edge-edge hybridization and effects of current conservation as discussed in Sec.~\ref{sec:current_density}. We also note that for large $\mathcal{R}$, the size of the positive and negative portions of $\vj(\vR)$ become more similar in size implying a smaller total integrated current $I_x$, while they are very dissimilar due to a larger (smaller) positive portion at $\mathcal{R} = 20\xi_0$ ($\mathcal{R} = 5\xi_0$) implying a large positive (negative) current. These effects are most pronounced at low temperatures, which is in agreement with the main text in Sec.~\ref{sec:total_current}.

Figure~\ref{fig:chiral_currents_kappa} shows the charge-current density versus distance from the edge in the large regime $\mathcal{R} > 20\xi_0$ (a), and the small regime $\mathcal{R} < 20\xi_0$ (b). Different lines correspond to different penetration depths $\lambda_0$. Above $\lambda_0>\mathcal{R}$, there is negligible effect of screening (small difference between finite and infinite $\lambda_0$). As $\lambda_0$ reduces below $\mathcal{R}$, the magnitude of the positive portion decreases, while the negative portion increases, which is seen most clearly in Fig.~\ref{fig:chiral_currents_kappa}(a). Hence, the total current reduces, consistent with a stronger diamagnetic Meissner screening from the bulk condensate. However, as $\lambda_0$ becomes comparable with $\xi_0$, a qualitatively different behavior develops. The current obtains a more complicated spatial dependence with kinks, a smaller negative region, and an additional sign change far from the edge in agreement with semi-infinite systems studied in Ref.~\cite{Wang:2018}. This leads to a modified balance for the integrated current, such that it changes sign as shown in the main text in Sec.~\ref{sec:chiral_ground_state:induction}.

\section{Fit of asymptotic behavior for large system size}
\label{app:fit}
In this appendix, we provide additional details on the fit procedure used in Sec.~\ref{sec:total_current} for the total charge current $I_x$ in Fig.~\ref{fig:total_current_radius_temperature}(c) and the OMM $m_z$ in Fig.~\ref{fig:magnetic_moment_radius_temperature}(c) as a function of $\mathcal{R}$.

We initially compare different fit functions, specifically exponential decay and various polynomial decays. We however do not find a good agreement with exponential decay and focus in the following on the polynomial function. We limit ourselves to only considering polynomials with a few terms to reduce over-fitting, using the fit function
\begin{align}
    \label{eq:fit_function:2}
    f_{\mathrm{fit}}(\mathcal{R}) = c_0 + c_1\mathcal{R}^{-|c_2|},
\end{align}
with fit parameters $c_0$, $c_1$, and $c_2$. We find that the fit function which most accurately describe the asymptotic behavior of $I_x$ and $m_z$ for large $\mathcal{R}$ to be the polynomial $1/\mathcal{R}$ (best fit with $c_2\approx 1 \pm 0.1$), while the higher-order polynomials (e.g. $1/\mathcal{R}^2$) result in very poor fits. Moreover, we verify that fitting only a few data points, e.g. $30\xi_0 \leq \mathcal{R} \leq 80\xi_0$ [gray region in Figs.~\ref{fig:total_current_radius_temperature}(c) and \ref{fig:magnetic_moment_radius_temperature}(c)] , gives a good extrapolation to all other data points in the interval $\mathcal{R} \geq 20\xi_0$.

Table~\ref{tab:fit_constants} summarizes the best fit parameters for $I_x$ and $m_z$ for fixed $c_2=1$.
\begin{table}[h!]
\def\arraystretch{1.3}
\begin{ruledtabular}
\begin{tabular}{c | c | c | c}
    $f_{\mathrm{fit}}$ & $T/\Tc$ & $c_0$ & $c_1$ \\
\hline
    $I_x$ & $0.02$ & $3\times10^{-4} I_0$ & $0.63$ \\
    $m_z$ & $0.02$ & -$5\times10^{-4} m_0$  & $0.78$ \\
\end{tabular}
\end{ruledtabular}
\caption{Best fit parameters $c_{0,1}$ for $I_x(\mathcal{R})$ and $m_z(\mathcal{R})$ with fit function $f_{\mathrm{fit}}(\mathcal{R})$ [Eq.~(\ref{eq:fit_function:2})], at different temperatures and fixed $c_2 = 1$.}
\label{tab:fit_constants}
\end{table}
From these fit parameters, we find that as $T\to0$ and $\mathcal{R}\to\infty$, both $I_x$ and $m_z$ tends to a very small, negligible, value. An exact zero is of course not found due to finite temperature $T/\Tc=0.02$ and finite numerical accuracy. The numerical accuracy causes some uncertainty in $c_0$ for each scenario depending on fit region, but $c_1$ remains more or less the same with $c_1 \approx 0.7 \pm 0.1$. The fact that both $I_x$ and $m_z$ have the same limiting behavior for large $\mathcal{R}$ is understood from their expressions in Eqs.~(\ref{eq:sheet_current}) and (\ref{eq:model:magnetic_moment}): the total current is the integral of $\vj(\vR)$ over the radius, while the magnetic moment is the integral over $\vR \times \vj(\vR)$ over the area.

\bibliographystyle{apsrev4-2}
\bibliography{cite.bib}

\end{document}